\documentclass[12pt]{article}

\usepackage{amssymb}
\usepackage{amsmath}

\usepackage{graphicx}

\usepackage{setspace}

\makeatletter
\newsavebox{\@brx}
\newcommand{\llangle}[1][]{\savebox{\@brx}{\(\m@th{#1\langle}\)}%
  \mathopen{\copy\@brx\kern-0.5\wd\@brx\usebox{\@brx}}}
\newcommand{\rrangle}[1][]{\savebox{\@brx}{\(\m@th{#1\rangle}\)}%
  \mathclose{\copy\@brx\kern-0.5\wd\@brx\usebox{\@brx}}}
\makeatother

\newcommand{\qed}{\hfill \rule{2.3mm}{2.3mm}}
\newcommand{\fr}[2]{\frac{\displaystyle #1}{\displaystyle #2}}
\newcommand{\df}[2]{\frac{\displaystyle d#1}{\displaystyle d#2}}

\newcommand{\ve}{\varepsilon}

\newcommand{\Real}{\mathrm{Re}\:}
\newcommand{\Imag}{\mathrm{Im}\:}

\newtheorem{theorem}{Theorem}[section]

\newtheorem{proposition}[theorem]{Proposition}
\newtheorem{remark}[theorem]{Remark}

\begin{document}
\title{Mathematical framework for breathing chimera states}

\author{O. E. Omel'chenko\\[3mm]
Institute of Physics and Astronomy, University of Potsdam,\\
Karl-Liebknecht-Str. 24/25, 14476 Potsdam, Germany,\\[2mm]
{\bf E-Mail:} omelchenko@uni-potsdam.de}

\maketitle



\begin{abstract}
About two decades ago it was discovered that systems of nonlocally coupled oscillators
can exhibit unusual symmetry-breaking patterns composed of coherent and incoherent regions.
Since then such patterns, called chimera states, have been the subject of intensive study
but mostly in the stationary case when the coarse-grained system dynamics remains unchanged over time.
Nonstationary coherence-incoherence patterns, in particular periodically breathing chimera states,
were also reported, however not investigated systematically because of their complexity.
In this paper we suggest a semi-analytic solution to the above problem
providing a mathematical framework for the analysis of breathing chimera states
in a ring of nonlocally coupled phase oscillators. Our approach relies on the consideration
of an integro-differential equation describing the long-term coarse-grained dynamics of the oscillator system.
For this equation we specify a class of solutions relevant to breathing chimera states.
We derive a self-consistency equation for these solutions and carry out their stability analysis.
We show that our approach correctly predicts macroscopic features of breathing chimera states.
Moreover, we point out its potential application to other models
which can be studied using the Ott-Antonsen reduction technique.
\end{abstract}

\section{Introduction}
\label{Sec:Intro}

Many living organisms, chemical and physical systems can behave as self-sustained oscillators~\cite{Book:Winfree}.
Spiking neurons in the brain, flashing fireflies, Belousov-Zhabotinsky (BZ) reaction
and swinging pendula of metronomes are just a few examples of such kind.
When two or more self-sustained oscillators interact with each other,
their rhythms tend to adjust in a certain order
resulting in their partial or complete synchronization~\cite{Book:PikovskyRK,AreD-GKMZ2008}.
Such phenomena have been observed in many real-world systems and laboratory experiments,
including Josephson junction arrays~\cite{WieCS1996}, populations of fireflies~\cite{BucB1968} and yeast cells~\cite{DeMODS2008},
chemical~\cite{TayTWHS2009} and electrochemical oscillators~\cite{KisZH2002}.
Moreover, synchronization turns out to underlie many physiological processes~\cite{Gla2001,YamIMOYKO2003}
and is, in some cases, associated with certain brain disorders,
such as schizophrenia, epilepsy, Alzheimer's and Parkinson's diseases~\cite{UhlS2006,LehBHKRSW2009}.

Mathematical modelling of synchronization processes often relies
on different versions of the Kuramoto model~\cite{Book:Kuramoto,Str2000,AceBVRS2005,RodPJK2016},
where the state of each oscillator is described by a single scalar variable,  its phase.
These models are derived as normal forms
for general weakly coupled oscillator networks~\cite{Book:HoppensteadtI,AshR2016,PieD2019}.
Moreover, the resulting equations are simplified additionally to reduce their complexity.
This paper is concerned with a Kuramoto model of the form
\begin{equation}
\df{\theta_k}{t} = - \frac{2 \pi}{N} \sum\limits_{j=1}^N
G\left( \fr{2\pi (k - j)}{N} \right) \sin( \theta_k(t) - \theta_j(t) + \alpha),\qquad k=1,\dots,N.
\label{Eq:Oscillators}
\end{equation}
Here~$G(x)$ is a nonconstant continuous even function called {\it coupling kernel}
and $\alpha\in(-\pi/2,\pi/2)$ is a phase lag parameter.
System~(\ref{Eq:Oscillators}) describes dynamics of~$N$ identical nonlocally coupled phase oscillators~$\theta_k$.
Moreover, if~$G(x)$ is $2\pi$-periodic, then the connections between oscillators have circular symmetry
and system~(\ref{Eq:Oscillators}) is in fact a ring of coupled oscillators.
Model~(\ref{Eq:Oscillators}) was first suggested by Kuramoto and Battogtokh in~\cite{KurB2002}
and since then has been intensively studied in the context of chimera states.
By chimera states one denotes specific dynamical regimes in system~(\ref{Eq:Oscillators})
where a part of oscillators get synchronized, while the others keep oscillating asynchronously.
Such states are irrelevant to the circular symmetry of system~(\ref{Eq:Oscillators}),
therefore their emergence seems to be counterintuitive,
what explains the origin of their name~\cite{AbrS2004}.
Being first reported for nonlocally coupled phase oscillators,
later chimera states have been found in a variety of other coupled oscillator systems,
both in experiments~\cite{HagMRHOS2012,KapKWCM2014,MarTFH2013,RosRHSG2014,SchSKG-M2014,TinNS2012,TotRTSE2018,WicK2013},
and in numerical simulations, see~\cite{PanA2015,KemHSKK2016,Sch2016,Ome2018,MajBGP2019} and references therein.

\begin{figure}[ht]
\begin{center}
\includegraphics[width=0.6\textwidth]{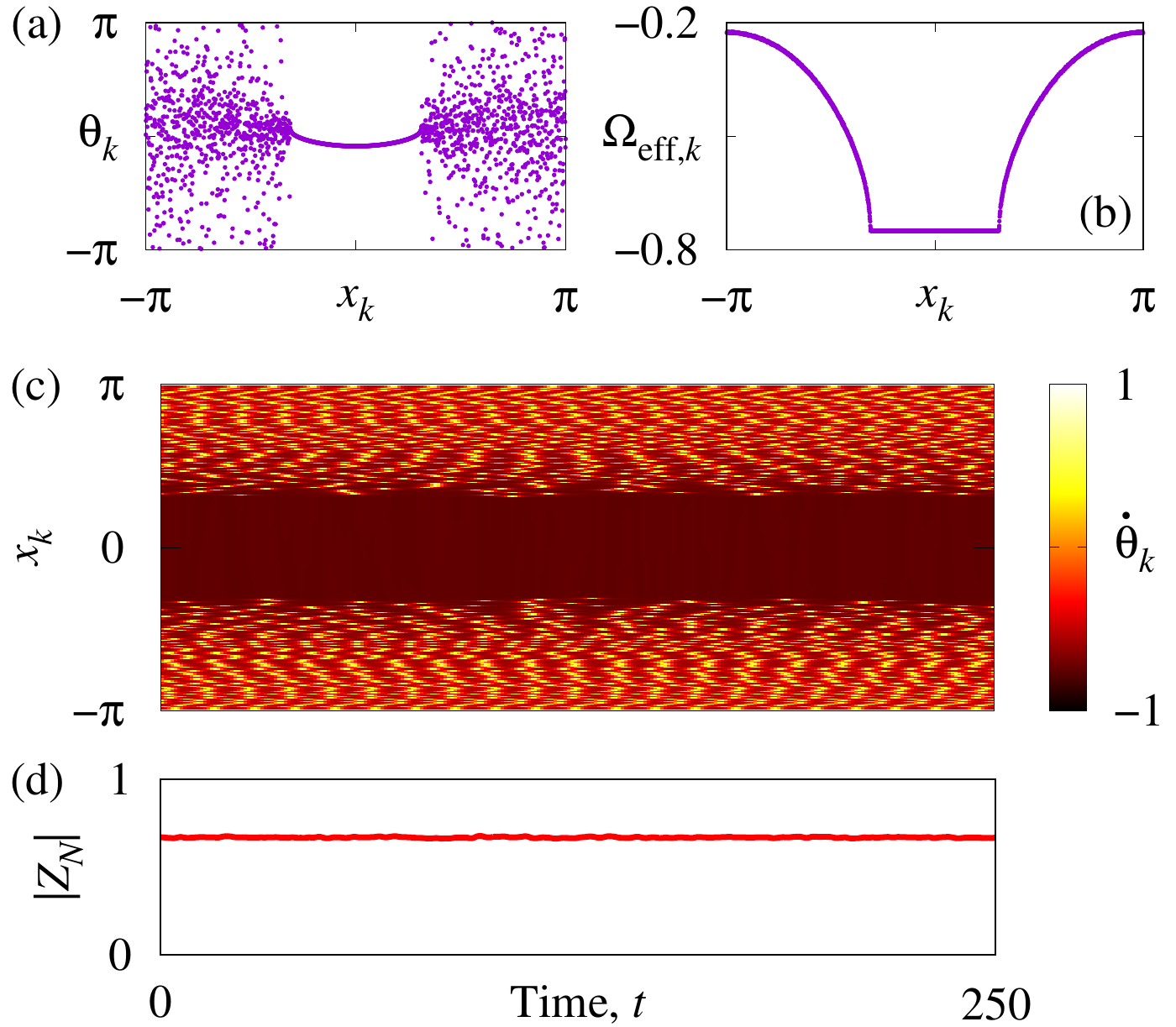}
\end{center}
\caption{
A stationary chimera state in the system~(\ref{Eq:Oscillators}) with cosine coupling kernel~(\ref{Coupling:Cos}).
(a) Snapshot. (b) Effective frequencies~$\Omega_{\mathrm{eff},k}$ as defined in~(\ref{Def:Omega_eff}).
(c) Space-time plot of phase velocities.
(d) Modulus of the global order parameter~$Z_N(t)$ given by~(\ref{Def:Z}).
Parameters: $N = 8192$, $A = 0.9$ and $\alpha = \pi/2 - 0.16$.
}
\label{Fig:P}
\end{figure}

A typical example of a chimera state in the system~(\ref{Eq:Oscillators}) with a cosine coupling kernel
\begin{equation}
G(x) = \fr{1}{2\pi} (1 + A \cos x),\qquad A>0,
\label{Coupling:Cos}
\end{equation}
is shown in Figure~\ref{Fig:P}.
Note that, for convenience, in the panels~(a)--(c) we use the oscillator positions $x_k = -\pi + 2\pi k / N$ instead of the discrete indices~$k$.
The main difference between synchronized and asynchronous oscillators can be seen in the space-time plot~(c).
The former oscillators rotate at almost the same constant speed,
while the latter oscillators have a more complicated dynamics.
As a result, in the snapshot~(a) there appear two regions:
a coherent region, where the phases~$\theta_k$ lie on a smooth curve,
and an incoherent region, where the phases are randomly distributed.
Moreover, if we compute the {\it effective frequencies} $\Omega_{\mathrm{eff},k}$ defined by
\begin{equation}
\Omega_{\mathrm{eff},k} = \lim\limits_{\tau\to\infty} \fr{1}{\tau} \int_0^\tau \df{\theta_k}{t} dt,
\label{Def:Omega_eff}
\end{equation}
then we obtain the arc-shaped graph~(b) with a plateau corresponding to synchronized oscillators.
Note that although the oscillator dynamics behind the chimera state is quite complicated and chaotic~\cite{WolOYM2011},
in the statistical sense, this state is stationary as can be seen from the graph~(d)
of the global order parameter
\begin{equation}
Z_N(t) = \fr{1}{N} \sum\limits_{k=1}^N e^{i \theta_k(t)}.
\label{Def:Z}
\end{equation}

In this paper, we consider a more complicated and therefore less explored type of chimera states,
called {\it breathing chimera states}, which is characterized by nonstationary macroscopic dynamics.
\begin{figure}[ht]
\begin{center}
\includegraphics[width=0.6\textwidth]{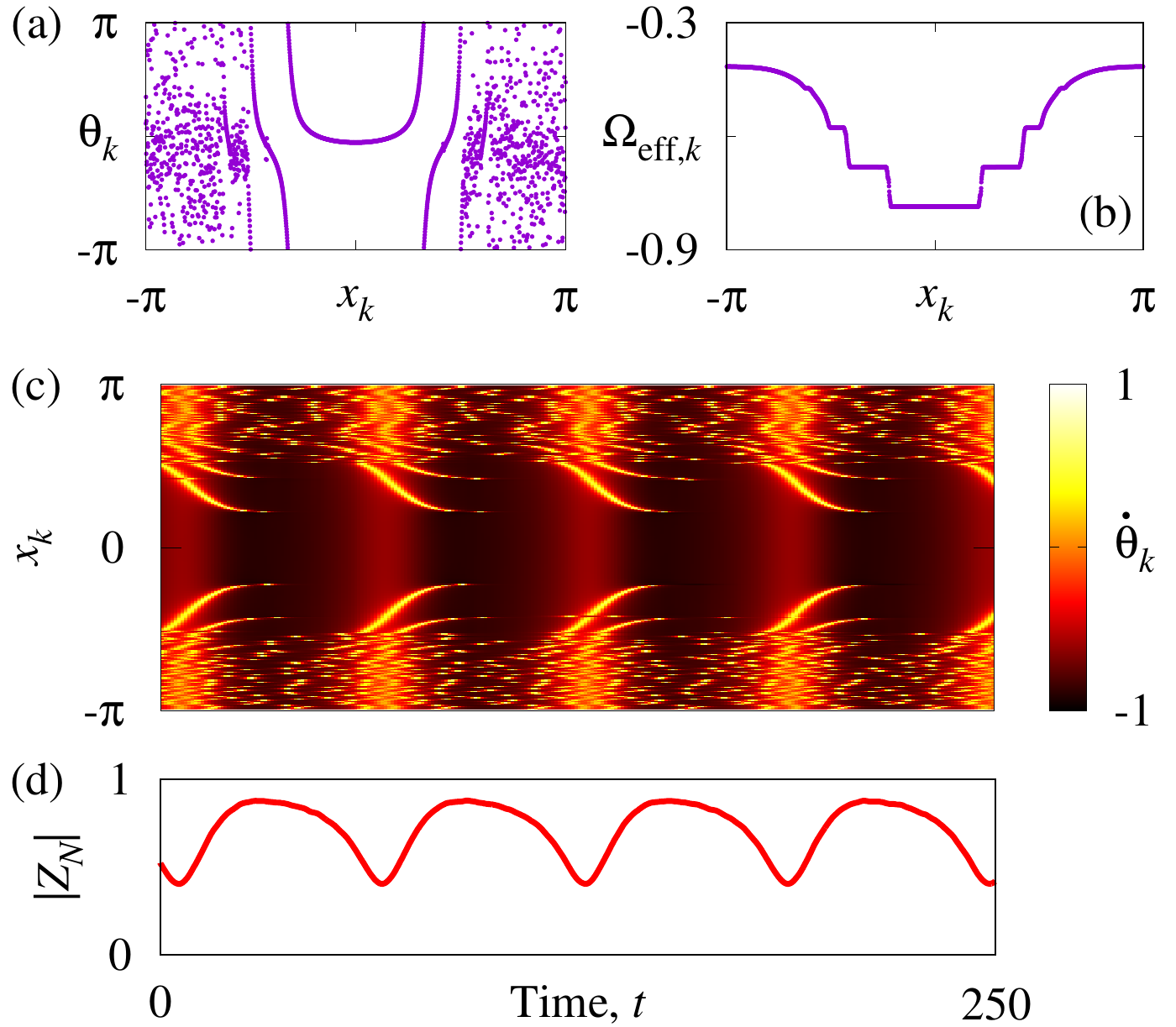}
\end{center}
\caption{A breathing chimera state in the system~(\ref{Eq:Oscillators}) with cosine coupling kernel~(\ref{Coupling:Cos}).
(a) Snapshot. (b) Effective frequencies~$\Omega_{\mathrm{eff},k}$ as defined in~(\ref{Def:Omega_eff}).
(c) Space-time plot of phase velocities.
(d) Modulus of the global order parameter~$Z_N(t)$ given by~(\ref{Def:Z}).
Parameters: $N = 8192$, $A = 1.05$ and $\alpha = \pi/2 - 0.16$.
}
\label{Fig:QP}
\end{figure}
An example of such state is shown in Figure~\ref{Fig:QP}.
Roughly speaking, the typical features of breathing chimera states are:

(i) multiple synchronized and asynchronous regions in the snapshot of~$\theta_k$,

(ii) multiple equidistant plateaus in the graph of the effective frequencies $\Omega_{\mathrm{eff},k}$,

(iii) periodically varying (breathing) oscillator dynamics,

(iv) oscillating modulus~$|Z_N(t)|$ of the global order parameter.

Apart from the above example breathing chimera states have been found in systems~(\ref{Eq:Oscillators})
with exponential~\cite{BolSOP2018} and top-hat~\cite{SudO2018,SudO2020} coupling kernels~$G(x)$.
They were also observed in two-dimensional lattices of phase oscillators
as breathing spirals with randomized cores~\cite{XieKK2015,OmeWK2018}.
Moreover, it was shown that breathing chimera states persist
for slightly heterogeneous phase oscillators~\cite{Ome2020a,Lai2009}
and can emerge in systems of coupled limit cycle oscillators~\cite{GopCVL2014,CleCFR2018}.
In spite of the number of these results the true nature of breathing chimera states still remains unclear.
Even for relatively simple system~(\ref{Eq:Oscillators}) no mathematical methods are known
to analyze their stability or predict their properties for given coupling kernel~$G(x)$ and phase lag~$\alpha$.
This is the problem we are going to address below.

Our approach is based on the consideration of an integro-differential equation
\begin{equation}
\frac{d z}{d t} = \frac{1}{2} e^{-i \alpha} \mathcal{G} z
- \frac{1}{2} e^{i \alpha} z^2 \mathcal{G} \overline{z},
\label{Eq:OA}
\end{equation}
where~$z(x,t)$ is an unknown complex-valued function $2\pi$-periodic with respect to~$x$,
$\overline{z}$~denotes the complex conjugate of~$z$
and symbol~$\mathcal{G}$ denotes an integral operator
\begin{equation}
\mathcal{G}\::\: C_\mathrm{per}([-\pi,\pi];\mathbb{C})\to C_\mathrm{per}([-\pi,\pi];\mathbb{C}),\qquad
( \mathcal{G} u)(x) = \int_{-\pi}^\pi G(x-y) u(y) dy
\label{Def:G}
\end{equation}
with the coupling kernel~$G(x)$ identical to that in system~(\ref{Eq:Oscillators}).
It is known~\cite{Ome2018,Ome2013} that in the limit of infinitely many oscillators $N\to\infty$,
Eq.~(\ref{Eq:OA}) describes the long-term coarse-grained dynamics of system~(\ref{Eq:Oscillators}).
If one assumes that $x_k = -\pi + 2\pi k / N$ is the physical position of the $k$th oscillator,
then $z(x,t)$~yields the local order parameter of the oscillators positioned around $x\in[-\pi,\pi]$.
More precisely, this means
$$
z(x,t) = \fr{1}{\# \{ k\::\: | x_k - x | < \delta \} } \sum\limits_{k\::\: |x_k - x|<\delta} e^{i \theta_k(t)},
$$
where $0 < \delta << 1$ and $\#\{\cdot\}$ denotes the number of indices~$k$ that satisfy the condition in curly brackets.
If $|z(x,t)| = 1$, then the oscillators with $x_k\approx x$ behave synchronously, in other words, they are coherent.
In contrast, if $|z(x,t)| < 1$, then the oscillators with $x_k\approx x$ behave asynchronously.
This interpretation implies that only functions~$z(x,t)$ satisfying the inequality $|z(x,t)|\le 1$
are relevant to the oscillator system~(\ref{Eq:Oscillators}).
Moreover, every chimera state is represented by a spatially structured solution~$z(x,t)$
composed of both coherent ($|z| = 1$) and incoherent ($|z| < 1$) regions.

In most of previous studies, chimera states were identified with the rotating wave solutions of Eq.~(\ref{Eq:OA}) given by
\begin{equation}
z = a(x) e^{i \Omega t},
\label{Ansatz:RW}
\end{equation}
where $a\in C_\mathrm{per}([-\pi,\pi];\mathbb{C})$ and $\Omega\in\mathbb{R}$.
However, for breathing chimera states ansatz~(\ref{Ansatz:RW}) does not work.
Numerical results in~\cite{SudO2020,OmeWK2018} suggest
that, in this case, one has to look for more complicated solutions
\begin{equation}
z = a(x,t) e^{i \Omega t},
\label{Ansatz:QP}
\end{equation}
where~$a(x,t)$ is a function $2\pi$-periodic with respect to~$x$
and $T$-periodic with respect to~$t$ for some~$T>0$.
To distinguish the cyclic frequencies~$\Omega$ and~$\omega = 2\pi/T$
in the following we call them {\it the primary frequency} and {\it the secondary frequency}, respectively.
In general these frequencies are different and incommensurable,
therefore the dynamics of the solution~(\ref{Ansatz:QP}) appears to be quasiperiodic,
thus breathing chimera states can also be called {\it quasiperiodic chimera states}.

\begin{remark}
Note that the product ansatz~(\ref{Ansatz:QP}) with a $T$-periodic function $a(x,t)$, in general, is not uniquely determined.
Indeed, for every nonzero integer $m$ we can rewrite it in the equivalent form
$$
z = a_m(x,t) e^{i \Omega_m t}\quad\mbox{with}\quad
a_m(x,t) = a(x,t) e^{i m \omega t}\quad\mbox{and}\quad \Omega_m = \Omega - m \omega.
$$
To avoid this ambiguity, throughout the paper we assume that the function $a(x,t)$
and the constant $\Omega$ in~(\ref{Ansatz:QP}) are chosen so that
$$
\lim\limits_{\tau\to\infty} \frac{1}{\tau} \int_0^\tau d \arg Y(t) = 0
\qquad\mbox{where}\qquad
Y(t) = \frac{1}{2\pi} \int_{-\pi}^\pi a(x,t) dx.
$$
Roughly speaking, we request that the variation of the complex argument of $Y(t)$
remains bounded for all $t \ge 0$.

Importantly, the above calibration condition is well-defined only if $Y(t)\ne 0$ for all $t\ge 0$,
therefore we checked carefully that this requirement is satisfied
for all examples of breathing chimera states shown in the paper.
\label{Remark:Calibration}
\end{remark}

\begin{remark}
The term quasiperiodic chimera state was previously used in~\cite{PikR2008}
to denote some partially synchronized states in a two population model
of globally coupled identical phase oscillators.
In contrast to breathing chimera states considered here,
the quasiperiodic chimera states from~\cite{PikR2008} are characterized
by quasiperiodically (not periodically!) oscillating modulus $|Z_N(t)|$ of the global order parameter.
Moreover, their mathematical description goes beyond the Ott-Antonsen theory~\cite{OttA2008} used in this paper
and requires the application of a more general approach suggested by Watanabe and Strogatz~\cite{WatS1993,WatS1994}.
The chimera state in Figure~\ref{Fig:QP} demonstrates a simpler, periodic dynamics of $|Z_N(t)|$,
therefore, in the context of two population models, it should be compared
with the breathing chimera states reported in~\cite{AbrMSW2008}.
\end{remark}

\begin{remark}
It is easy to verify that Eq.~(\ref{Eq:OA}) is equivariant with respect
to the one-parameter Lie group of complex phase shifts $z \mapsto z e^{i \phi}$ with $\phi\in\mathbb{R}\backslash 2\pi\mathbb{Z}$.
Therefore, in the mathematical literature, solutions of Eq.~(\ref{Eq:OA}) of the form~(\ref{Ansatz:RW}) and~(\ref{Ansatz:QP})
are called relative equilibria and relative periodic orbits (with respect to this group action), respectively.
These names also indicate that solution~(\ref{Ansatz:RW}) becomes an equilibrium
and solution~(\ref{Ansatz:QP}) becomes a periodic orbit in the appropriate corotating frame.
The existence and stability of relative equilibria of Eq.~(\ref{Eq:OA})
were in the focus of many papers, in particular, those concerned with stationary chimera states.
In contrast, relative periodic orbits of Eq.~(\ref{Eq:OA}) were hardly considered.
The main purpose of the present paper is to fill this gap.
\end{remark}

In this paper, we demonstrate that ansatz~(\ref{Ansatz:QP}) does describe breathing chimera states
and can be used for their continuation as well as for their stability analysis.
The paper is organized as follows.
In Section~\ref{Sec:Riccati} we consider an auxiliary complex Riccati equation
$$
\df{u}{t} = w(t) - i s u(t) - \overline{w}(t) u^2(t),
$$
where~$s$ is a real coefficient and~$w(t)$ is a continuous complex-valued function.
We show that, in general, this equation has a unique stable periodic solution satisfying the inequality~$|u(t)|\le 1$.
The corresponding solution operator is denoted by
$$
\mathcal{U}\::\: (w,s) \in C_\mathrm{per}([0,2\pi];\mathbb{C}) \times \mathbb{R} \mapsto u\in C^1_\mathrm{per}([0,2\pi];\mathbb{C}).
$$
Its properties are discussed in Sections~\ref{Sec:Operator:U} and~\ref{Sec:Operator:U:Derivatives}.
Although the operator~$\mathcal{U}$ is defined implicitly, it turns out that its value can be computed
by solving only three initial value problems for the above complex Riccati equation.

In Section~\ref{Sec:SC}, we show that if Eq.~(\ref{Eq:OA}) has a stable solution of the form~(\ref{Ansatz:QP}),
then its amplitude~$a(x,t)$ and its primary and secondary frequencies~$\Omega$ and~$\omega$
satisfy a self-consistency equation
\begin{equation}
2 \omega e^{i \alpha} w(x,t) - \mathcal{G} \mathcal{U}( w(x,t), s ) = 0,
\label{Eq:SC}
\end{equation}
where
\begin{equation}
\omega = \fr{2\pi}{T},\qquad
s = \fr{\Omega}{\omega},\qquad
w(x,t) = \fr{e^{-i \alpha}}{2 \omega} \mathcal{G} a\left( x, \fr{t}{\omega} \right).
\label{Def:omega:s:w}
\end{equation}
A modified version of Eq.~(\ref{Eq:SC}) is obtained in Section~\ref{Eq:SC:Modified}.
Then in Section~\ref{Eq:SC:Modified:Cos} we suggest a continuation algorithm
allowing to compute the solution branches of Eq.~(\ref{Eq:SC})
and thus to predict the properties of breathing chimera states.
In particular, in Section~\ref{Sec:Formulas:Z:Omega_eff}
we explain how the continuum limit analogs of the global order parameter~$Z_N(t)$
and of the effective frequencies~$\Omega_{\mathrm{eff},k}$ can be computed.
Moreover, for the sake of completeness, in Section~\ref{Sec:Extraction}
we also explain how to extract the primary and secondary frequencies
of a breathing chimera state from the corresponding trajectory of coupled oscillator system~(\ref{Eq:Oscillators}).

In Section~\ref{Sec:Stability}, we carry out the linear stability analysis of a general solution~(\ref{Ansatz:QP}) to Eq.~(\ref{Eq:OA}).
It relies on the consideration of the monodromy operator,
which describes the evolution of small perturbations in the linearized Eq.~(\ref{Eq:OA}).
More precisely, the stability of the solution~(\ref{Ansatz:QP}) is determined by the spectrum of the monodromy operator.
The spectrum consists of two parts: essential and discrete spectra.
The former part is known explicitly and has no influence on the stability of breathing chimera states,
while the latter part is crucial for their stability but can be computed only numerically as explained in Section~\ref{Sec:DiscreteSpectrum}.

In Section~\ref{Sec:Example}, we illustrate the performance of the developed methods
considering specific examples of breathing chimera states in system~(\ref{Eq:Oscillators}).
In particular, using the continuation method, we find a branch of breathing chimera states
starting from the chimera state in Fig.~\ref{Fig:QP}.
Moreover, we also compute theoretical predictions of the graphs of~$|Z_N(t)|$ and~$\Omega_{\mathrm{eff},k}$
shown in Fig.~\ref{Fig:QP}(b),(d).
Furthermore, our analysis reveals that solutions~(\ref{Ansatz:QP}) can lose their stability
only via nonclassical bifurcations when one or two unstable eigenvalues emerge from the essential spectrum.
Similar results were known for partially synchronized states in the classical Kuramoto model~\cite{MirS2007}
and for stationary chimera states~(\ref{Ansatz:RW})~\cite{Ome2013}.
But for breathing chimera states they are reported for the first time here.
Finally, in Section~\ref{Sec:Conclusions} we summarize the obtained results
and point out other potential applications of our methods.

{\bf Notations.} Throughout this paper we use the following notations.
We let $C_\mathrm{per}([-\pi,\pi];\mathbb{C})$ denote
the space of all $2\pi$-periodic continuous complex-valued functions.
A similar notation $C_\mathrm{per}([-\pi,\pi]\times[0,2\pi];\mathbb{C})$
is used to denote the space of all continuous double-periodic functions
on the square domain $[-\pi,\pi]\times[0,2\pi]$.
Moreover, the capital calligraphic letters such as~$\mathcal{G}$ or~$\mathcal{U}$
are used to denote operators on appropriate Banach spaces.
Finally, the symbol $\Theta(t)$ denotes the Heaviside step function
such that $\Theta(t) = 0$ for $t < 0$ and $\Theta(t) = 1$ for $t \ge 0$.

\section{Periodic complex Riccati equation}
\label{Sec:Riccati}

Let us consider a complex Riccati equation of the form
\begin{equation}
\df{u}{t} = w(t) - i s u(t) - \overline{w}(t) u^2(t),
\label{Eq:Riccati}
\end{equation}
where $s\in\mathbb{R}$ and~$w(t)$ is a continuous complex-valued function.
Let $\mathbb{D} = \{ v\in\mathbb{C}\::\:|v| < 1 \}$ denote the open unit disc of the complex plane
and $\overline{\mathbb{D}} = \mathbb{D}\cup\partial \mathbb{D}$ be its closure.
In this section we will show that for every $(w(t),s)\in C_\mathrm{per}([0,2\pi];\mathbb{C})\times\mathbb{R}$
such that $|s| + \max\limits_{t\in[0,2\pi]} |w(x)| \ne 0$, in general,
there exists a unique stable solution to Eq.~(\ref{Eq:Riccati})
lying entirely in the unit disc~$\overline{\mathbb{D}}$.
The nonlinear operator yielding this solution will be denoted by $\mathcal{U}(w(t),s)$.

\begin{proposition}
For every $s\in\mathbb{R}$, $w\in C(\mathbb{R};\mathbb{C})$ and~$u_0 \in \overline{\mathbb{D}}$
there exists a unique global solution to equation~(\ref{Eq:Riccati})
starting from the initial condition $u(0) = u_0$.
Moreover, if $|u_0| = 1$ or $|u_0| < 1$,
then $|u(t)| = 1$ or $|u(t)| < 1$ for all~$t\in\mathbb{R}$, respectively.
\label{Proposition:Disc}
\end{proposition}

{\bf Proof:}
Suppose that~$u(t)$ is a solution to equation~(\ref{Eq:Riccati}), then
\begin{eqnarray}
\df{|u|^2}{t} &=& u(t) \df{\overline{u}}{t} + \overline{u}(t) \df{u}{t}
= u(t) \overline{w}(t) + i s |u(t)|^2 - w(t) |u(t)|^2 \overline{u}(t) \nonumber\\[2mm]
&+& \overline{u}(t) w(t) - i s |u(t)|^2 - \overline{w}(t) |u(t)|^2 u(t)
= 2 \mathrm{Re}( u(t) \overline{w}(t) ) ( 1 - |u(t)|^2 ).
\label{Eq:Modulus:u}
\end{eqnarray}
According to Eq.~(\ref{Eq:Modulus:u}), if $|u(0)| = 1$ then $|u(t)| = 1$ for all other~$t\ne 0$,
therefore the solution~$u(t)$ cannot blow up in finite time
and hence it can be extended for all $t\in\mathbb{R}$.
On the other hand, Eq.~(\ref{Eq:Modulus:u}) implies that every solution~$u(t)$
satisfying $|u(0)| < 1$ remains trapped inside the disc~$\mathbb{D}$,
therefore it also can be extended for all $t\in\mathbb{R}$.~\qed

\begin{remark}
Every solution~$u(t)$ to Eq.~(\ref{Eq:Riccati}) satisfying the identity $|u(t)| = 1$
can be written in the form $u(t) = e^{i \psi(t)}$
where $\psi(t)$ is a solution to the equation
$$
\df{\psi}{t} = - s + 2 \mathrm{Im}( w(t) e^{-i \psi} ).
$$
\label{Remark:u1}
\end{remark}

Now we consider Eq.~(\ref{Eq:Riccati}) with a $2\pi$-periodic coefficient~$w(t)$.
It is well-known~\cite{Cam1997,Wil2008} that the Poincar{\'e} map of such equation
coincides with the M{\"o}bius transformation.
Because of Proposition~\ref{Proposition:Disc} this M{\"o}bius transformation
maps unit disc~$\overline{\mathbb{D}}$ onto itself, therefore it can be written in the form
\begin{equation}
\mathcal{M}(u) = \fr{e^{i\theta}(u + b)}{\overline{b} u + 1}
\quad\mbox{where}\quad \theta\in\mathbb{R}\quad\mbox{and}\quad b\in\mathbb{C}.
\label{Def:M}
\end{equation}

\begin{remark}
The fact that the Poincar{\'e} map of the periodic complex Riccati equation~(\ref{Eq:Riccati})
coincides with the M{\"o}bius transformation~(\ref{Def:M}) can also be justified in a different way,
using the Lie theory, see~\cite[Sec.~III]{MarMS2009}.
\end{remark}

The next proposition shows that the parameters~$\theta$ and~$b$ in formula~(\ref{Def:M})
can be uniquely determined using two solutions to Eq.~(\ref{Eq:Riccati})
starting from the initial conditions $u = 0$ and $u = 1$.

\begin{proposition}
Suppose $s\in\mathbb{R}$ and $w\in C_\mathrm{per}([0,2\pi];\mathbb{C})$.
Let $U(t)$ and $\Psi(t)$ be solutions of the initial value problems
\begin{eqnarray}
\df{U}{t} &=& w(t) - i s U(t) - \overline{w}(t) U^2(t),\qquad U(0) = 0,
\label{IVP:U}\\[2mm]
\df{\Psi}{t} &=& - s + 2 \mathrm{Im}( w(t) e^{-i \Psi} ),\qquad\phantom{aaaaa} \Psi(0) = 0,
\label{IVP:psi}
\end{eqnarray}
and let $\zeta = U(-2\pi)$ and $\chi = \Psi(2\pi)$,
then the Poincar\'e map of Eq.~(\ref{Eq:Riccati}) is determined by the formula~(\ref{Def:M}) with
\begin{equation}
b = -\zeta
\qquad\mbox{and}\qquad
e^{i \theta} = \fr{ \overline{\zeta} - 1 }{\zeta - 1} e^{i\chi}.
\label{M:Params}
\end{equation}
Moreover $|b| < 1$.
\label{Proposition:U:psi}
\end{proposition}

{\bf Proof:} The definition of the Poincar{\'e} map and Remark~\ref{Remark:u1} imply
$$
\fr{e^{i\theta}(\zeta + b)}{\overline{b} \zeta + 1} = 0\qquad\mbox{and}\qquad \fr{e^{i\theta}(1 + b)}{\overline{b} + 1} = e^{i \chi}.
$$
The former equation yields $b = -\zeta$.
Inserting this into the latter equation we obtain a formula for $e^{i\theta}$.
Notice that because of Proposition~\ref{Proposition:Disc} we always have $|\zeta| < 1$, and hence $|b| < 1$ too.~\qed

Every $2\pi$-periodic solution to Eq.~(\ref{Eq:Riccati})
corresponds to a fixed point of the Poincar{\'e} map,
or equivalently to a solution of the equation
\begin{equation}
\mathcal{M}(u) = \fr{e^{i\theta}(u + b)}{\overline{b} u + 1} = u.
\label{Eq:FP}
\end{equation}
The periodic solution is stable or unstable,
if the corresponding fixed point~$u_*$ is stable or unstable
with respect to the map~$\mathcal{M}(u)$,
and the latter condition can be easily verified by estimating the derivative~$\mathcal{M}'(u_*)$.
Indeed, if $|\mathcal{M}'(u_*)| < 1$, then the fixed point~$u_*$ is stable.
On the other hand, if $|\mathcal{M}'(u_*)| > 1$, then~$u_*$ is unstable.
Moreover, the special properties of the map~$\mathcal{M}(u)$, see Remark~\ref{Remark:M:1}, allow us to conclude
that a fixed point~$u_*$ with $|\mathcal{M}'(u_*)| = 1$ is also stable provided it is non-degenerate.

In the next proposition, we show that every Poincar{\'e} map~(\ref{Def:M}) with $0<|b|<1$
has either a unique stable fixed point in the closed unit disc~$\overline{\mathbb{D}}$,
or a unique fixed point at all
(in this case, the fixed point is degenerate and lies on the unit disc boundary~$\partial\mathbb{D}$).

\begin{proposition}
Suppose $\theta\in(-\pi,\pi]$ and $0 < |b| < 1$, then Eq.~(\ref{Eq:FP})
has a unique solution~$u_0\in\overline{\mathbb{D}}$ such that $|\mathcal{M}'(u_0)| \le 1$.
This solution is given by the formulas
\begin{eqnarray}
&&
u_0 = \fr{ i \sin(\theta/2) + \sqrt{ |b|^2 - \sin^2(\theta/2) } }{|b|^2} b e^{i \theta / 2}\qquad\mbox{for}\qquad
|b| > |\sin (\theta/2)|,\label{Formula:u0:1}\\[2mm]
&&
u_0 = \fr{ i \sin(\theta/2) - i \sqrt{ \sin^2(\theta/2) - |b|^2 } }{|b|^2} b e^{i \theta / 2}\qquad\mbox{for}\qquad
|b| \le \sin (\theta/2),\nonumber\\[2mm]
&&
u_0 = \fr{ i \sin(\theta/2) + i \sqrt{ \sin^2(\theta/2) - |b|^2 } }{|b|^2} b e^{i \theta / 2}\qquad\mbox{for}\qquad
|b| \le -\sin (\theta/2).
\nonumber
\end{eqnarray}
Moreover, $|u_0| = 1$ for $|b| \ge |\sin (\theta/2)|$,
while $|u_0| < 1$ for $|b| < |\sin (\theta/2)|$.
Furthermore, $|\mathcal{M}'(u_0)| < 1$ for $|b| > |\sin (\theta/2)|$,
$|\mathcal{M}'(u_0)| = 1$ for $|b| < |\sin (\theta/2)|$,
and $\mathcal{M}'(u_0) = 1$ for $|b| = |\sin (\theta/2)|$.
\label{Proposition:IV}
\end{proposition}

{\bf Proof:} For every $|b| < 1$ and $u\in\overline{\mathbb{D}}$ equation~(\ref{Eq:FP}) can be rewritten in the form
\begin{equation}
e^{-i \theta/2} \overline{b} u^2 - 2 i \sin(\theta/2) u - e^{i \theta/2} b = 0.
\label{Eq:FP_}
\end{equation}
Since $b\ne 0$ this is a quadratic equation which generically has two complex roots.
We are going to check which of these roots lie in the unit disc~$\overline{\mathbb{D}}$
and what are their stability properties. To address the latter question
we compute the derivative of the M{\"o}bius transformation~(\ref{Def:M})
\begin{equation}
\mathcal{M}'(u) = \fr{e^{i \theta} ( 1 - |b|^2 )}{(\overline{b} u + 1)^2}
\label{Def:M:prime}
\end{equation}
and evaluate its modulus (keeping in mind that $|b| < 1$)
\begin{equation}
|\mathcal{M}'(u)| = \fr{1 - |b|^2}{|\overline{b} u + 1|^2}.
\label{M:prime}
\end{equation}

Depending on the sign of the difference $|b|^2 - \sin^2(\theta/2)$ we distinguish two cases.

{\it Case~1.} Suppose $|b| > |\sin(\theta/2)|$, then two solutions to Eq.~(\ref{Eq:FP_}) read
$$
u_\pm =  \fr{ i \sin(\theta/2) \pm \sqrt{ |b|^2 - \sin^2(\theta/2) } }{\overline{b} e^{-i \theta / 2}}.
$$
It is easy to verify that in this case $|u_+| = |u_-| = 1$. Moreover, we also obtain
\begin{eqnarray*}
1 - |b|^2 - |\overline{b} u_\pm + 1|^2 &=& 1 - |b|^2 - \left( \cos(\theta/2) \pm \sqrt{ |b|^2 - \sin^2(\theta/2) } \right)^2 \\[2mm]
&=& - 2 \sqrt{ |b|^2 - \sin^2(\theta/2) } \left( \sqrt{ |b|^2 - \sin^2(\theta/2) } \pm \cos(\theta/2) \right).
\end{eqnarray*}
Obviously, for every $\theta\in(-\pi,\pi]$ and $|\sin(\theta/2)| < |b| < 1$ we have
$$
\cos(\theta/2) = \sqrt{ 1 - \sin^2(\theta/2) } > \sqrt{ |b|^2 - \sin^2(\theta/2) },
$$
therefore
\begin{equation}
1 - |b|^2 - |\overline{b} u_+ + 1|^2 < 0\quad\mbox{and}\quad 1 - |b|^2 - |\overline{b} u_- + 1|^2 > 0,
\label{Ineq:u_pm}
\end{equation}
and hence
$$
| \mathcal{M}'(u_+) | < 1\quad\mbox{and}\quad | \mathcal{M}'(u_-) | > 1.
$$

{\it Case~2.} The other case is determined by the inequality $|b| \le |\sin(\theta/2)|$.
If $|b| < |\sin(\theta/2)|$, then Eq.~(\ref{Eq:FP_}) has two solutions
\begin{equation}
u_\pm =  \fr{ i \sin(\theta/2) \pm i \sqrt{ \sin^2(\theta/2) - |b|^2 } }{\overline{b} e^{-i \theta / 2}},
\label{Def:u:pm}
\end{equation}
while for $|b| = |\sin(\theta/2)|$ the values~$u_+$ and~$u_-$ given by~(\ref{Def:u:pm}) coincide.
To estimate the moduli~$|u_+|$ and~$|u_-|$ we compute the difference
$$
\left(  \sin(\theta/2) \pm \sqrt{ \sin^2(\theta/2) - |b|^2 } \right)^2 - |b|^2 = 2 \sqrt{ \sin^2(\theta/2) - |b|^2 } \left( \sqrt{ \sin^2(\theta/2) - |b|^2 } \pm \sin(\theta/2) \right).
$$
Then for $\sin(\theta/2) > 0$ we obtain
$$
\left(  \sin(\theta/2) + \sqrt{ \sin^2(\theta/2) - |b|^2 } \right)^2 - |b|^2 > 0,\quad
\left(  \sin(\theta/2) - \sqrt{ \sin^2(\theta/2) - |b|^2 } \right)^2 - |b|^2 < 0,
$$
and hence $|u_+| > 1$ and $|u_-| < 1$. Similarly, for $\sin(\theta/2) < 0$ we obtain $|u_+| < 1$ and $|u_-| > 1$.

Finally, we compute a difference relevant to formula~(\ref{M:prime})
$$
1 - |b|^2 - |\overline{b} u_\pm + 1|^2 = 1 - |b|^2 - \left| \cos(\theta/2) \pm i \sqrt{ \sin^2(\theta/2) -  |b|^2 } \right|^2 = 0,
$$
which implies $| \mathcal{M}'(u_+) | = | \mathcal{M}'(u_-) | = 1$.

On the other hand, in the limiting case $|b| = |\sin(\theta/2)|$, formulas~(\ref{Def:M:prime}) and~(\ref{Def:u:pm})
yield $|u_+| = |u_-| = 1$ and $\mathcal{M}'(u_+) = \mathcal{M}'(u_-) = 1$.~\qed

\begin{remark}
Let us consider the formula~(\ref{Def:M}) with $0 < |b| < |\sin(\theta/2)|$ in more detail.
In this case, $\mathcal{M}(u)$ determines an elliptic M{\"o}bius transformation, see \cite[Ch.~3.VII]{Needham}.
This means that it has two distinct fixed points that are neither attractive nor repulsive but indifferent.
(Recall the equation $| \mathcal{M}'(u_\pm) | = 1$ from the Case~2 in the proof of Proposition~\ref{Proposition:IV}.)
Moreover, the transformation moves all other points of the complex plane in circles around the two fixed points.
Therefore, according to the Lyapunov stability classification, both fixed points are stable, but not asymptotically stable.

Similarly one can verify that the other two cases $|\sin(\theta/2)| < |b| < 1$ and $|b| = |\sin(\theta/2)| \ne 0$
considered in Proposition~\ref{Proposition:IV} correspond
to the M{\"o}bius transformations~$\mathcal{M}(u)$ of hyperbolic and parabolic types, respectively.
This is in accordance with the fact that~$\mathcal{M}(u)$ has a pair of attracting and repulsive fixed points in the former case
and a degenerate fixed point in the latter case. Note that the degenerate fixed point
of a parabolic M{\"o}bius transformation is always unstable in the sense of Lyapunov~\cite[Ch.~3.VII]{Needham}.
\label{Remark:M:1}
\end{remark}

\begin{remark}
If $b = 0$, then Eq.~(\ref{Eq:FP}) degenerates into the linear equation $e^{i \theta} u = u$.
For $e^{i \theta} \ne 1$ this equation has only one solution $u = 0$,
while for $e^{i \theta} = 1$ it becomes trivial identity $u = u$
and hence has infinitely many solutions $u\in\overline{\mathbb{D}}$.
In both cases all the solutions are stable, because $\mathcal{M}(u)$ is linear and $| \mathcal{M}'(u) | = 1$.
Moreover, the case $b = 0$ and $e^{i \theta} = 1$ corresponds
to the equation~(\ref{Eq:Riccati}) with $w(t) = 0$ and $s = 0$.
\label{Remark:b:zero}
\end{remark}

\begin{remark}
If~$u_0$ is determined by formula~(\ref{Formula:u0:1}), then~$\mathcal{M}'(u_0)$ is real and $\mathcal{M}'(u_0)\in(0,1)$.
Indeed, formula~(\ref{Formula:u0:1}) implies
$$
\overline{b} u_0 =  \left( i \sin(\theta/2) + \sqrt{ |b|^2 - \sin^2(\theta/2) } \right) e^{i \theta / 2},
$$
therefore
$$
\overline{b} u_0 + 1 = \left( \cos(\theta/2) + \sqrt{ |b|^2 - \sin^2(\theta/2) } \right) e^{i \theta / 2}.
$$
Hence the assertion follows from formula~(\ref{Def:M:prime}) and from the first of two inequalities~(\ref{Ineq:u_pm}).
\label{Remark:M:real}
\end{remark}

\subsection{Solution operator~$\mathcal{U}$}
\label{Sec:Operator:U}

In the previous section we showed that for every~$w\in C_\mathrm{per}([0,2\pi];\mathbb{C})$ and $s\in\mathbb{R}$
the complex Riccati equation~(\ref{Eq:Riccati}) has a uniquely determined $2\pi$-periodic solution~$u(t)\in\overline{\mathbb{D}}$
that is stable in the sense of Lyapunov (or at least linearly stable in the degenerate case).
Let us denote the corresponding solution operator
$$
\mathcal{U}\::\: C_\mathrm{per}([0,2\pi];\mathbb{C})\times\mathbb{R}\to C_\mathrm{per}^1([0,2\pi];\mathbb{C}).
$$
The definition of~$\mathcal{U}$ is constructive and relies on the following steps:
\smallskip

1) Given~$w(t)$ and~$s$ one solves two initial value problems~(\ref{IVP:U}) and~(\ref{IVP:psi})
and obtains coefficients~$b$ and~$\theta$ of the M{\"o}bius transformation~(\ref{Def:M}), see Proposition~\ref{Proposition:U:psi}.
\smallskip

2) Using Proposition~\ref{Proposition:IV} one computes the initial value~$u_0$ of the periodic solution~$u(t)$
that lies entirely in the unit disc~$\overline{\mathbb{D}}$ and, moreover, is stable provided $|b| \ne |\sin(\theta/2)|$.
In the case $b = 0$, one assumes $u_0 = 0$, see Remark~\ref{Remark:b:zero}.
\smallskip

3) One integrates Eq.~(\ref{Eq:Riccati}) with the initial condition $u(0) = u_0$ and obtains $2\pi$-periodic solution~$u(t)$.
\smallskip

\noindent
Importantly, Propositions~\ref{Proposition:U:psi} and~\ref{Proposition:IV} ensure that the steps~1--3 can always be realized.
Therefore, the mapping $\mathcal{U}\::\: (w(t),s) \mapsto u(t)$ is well-defined.

\begin{remark}
Note that the minimal period of the function $u(t) = \mathcal{U}(w(t),s)$ does not have to be~$2\pi$.
In general, it can assume any value $2\pi/k$ with $k\in\mathbb{N}$.
Moreover, for certain values of the arguments $(w(t),s)$ the operator~$\mathcal{U}$ can also return a constant function~$u(t)$.
\end{remark}

Due to the definition of~$\mathcal{U}$ we have $|u(t)| \le 1$ for all $t\in[0,2\pi]$, therefore~$\mathcal{U}$ is a bounded operator.
Moreover, the operator~$\mathcal{U}$ has a specific dichotomy property:

\begin{proposition}
Let $w_*\in C_\mathrm{per}([0,2\pi];\mathbb{C})$, $s_*\in\mathbb{R}$ and $u_* = \mathcal{U}( w_*, s_* )$.
Moreover, let
\begin{equation}
M_* = \exp\left( - \int_0^{2\pi} ( i s_* + 2 \overline{w}_*(t) u_*(t) ) dt \right).
\label{Def:M_star}
\end{equation}
Then either $|u_*(t)| = 1$ for all $t\in[0,2\pi]$ and~$M_*$ is a real number such that $M_*\in(0,1]$,
or $|u_*(t)| < 1$ for all $t\in[0,2\pi]$ and $|M_*| = 1$.
\label{Proposition:Dichotomy}
\end{proposition}

{\bf Proof:} We need only to show that $M_* = \mathcal{M}'(u_0)$ where $u_0 = u_*(0)$.
Then the assertion follows from Propositions~\ref{Proposition:Disc} and~\ref{Proposition:IV}
and from Remarks~\ref{Remark:b:zero} and~\ref{Remark:M:real}.

Let us consider Eq.~(\ref{Eq:Riccati}) for $w(t) = w_*(t)$ and $s = s_*$.
Inserting there ansatz $u(t) = u_*(t) + v(t)$ and linearizing the resulting equation
with respect to small perturbations~$v(t)$ we obtain
\begin{equation}
\df{v}{t} = - ( i s_* + 2 \overline{w}_*(t) u_*(t) ) v.
\label{Eq:Linear:v}
\end{equation}
Obviously, formula~(\ref{Def:M_star}) determines the Floquet multiplier of the scalar linear equation~(\ref{Eq:Linear:v}).
By definition its value coincides with the derivative of the Poincar{\'e} map of the original nonlinear equation~(\ref{Eq:Riccati}),
hence $M_* = \mathcal{M}'(u_0)$ where $u_0 = u_*(0)$.~\qed

\begin{remark}
Proposition~\ref{Proposition:IV} and Remark~\ref{Remark:b:zero} imply
that $u_0 = u_*(0)$ is a simple fixed point of Eq.~(\ref{Eq:FP}), if and only if $M_* \ne 1$.
Therefore, the equation $M_* = 1$ can be considered as a degeneracy or bifurcation condition.
\end{remark}

\subsection{Derivatives of the operator~$\mathcal{U}$}
\label{Sec:Operator:U:Derivatives}

In this section we show how to compute partial derivatives of the operator~$\mathcal{U}$.

\begin{proposition}
Let $w_*\in C_\mathrm{per}([0,2\pi];\mathbb{C})$, $s_*\in\mathbb{R}$ and $u_* = \mathcal{U}( w_*, s_* )$.
Suppose
$$
\Phi_*(2\pi) \ne 1\qquad\mbox{where}\qquad
\Phi_*(t) = \exp\left( - \int_0^t ( i s_* + 2 \overline{w}_*(\tau) u_*(\tau) ) d\tau \right),
$$
then there exists a bounded linear operator $\mathcal{J}\::\: C_\mathrm{per}([0,2\pi];\mathbb{C}) \to C^1_\mathrm{per}([0,2\pi];\mathbb{C})$
given by
$$
(\mathcal{J} f)(t) = \int_0^{2\pi} \fr{\Phi_*(2\pi) + (1 - \Phi_*(2\pi)) \Theta(t - \tau)}{1 - \Phi_*(2\pi)} \Phi_*(t) \Phi_*^{-1}(\tau) f(\tau) d\tau
$$
such that $v(t) = (\mathcal{J} f)(t)$ is a $2\pi$-periodic solution to the equation
$$
\df{v}{t} + ( i s_* + 2 \overline{w}_*(t) u_*(t) ) v(t) = f(t).
$$
\label{Proposition:Green}
\end{proposition}

{\bf Proof:} This assertion has been proved in~\cite[Proposition~A.1]{Ome2019}.~\qed

\begin{proposition}
Let the assumptions of Proposition~\ref{Proposition:Green} be fulfilled.
Then for every $w\in C_\mathrm{per}([0,2\pi];\mathbb{C})$ we have
\begin{eqnarray}
\left. \partial_\ve \mathcal{U}( w_* + \ve w, s_* )\right|_{\ve = 0} &=& \mathcal{J} ( w - u_*^2 \overline{w} ),\label{Eq:U_w}\\[2mm]
\partial_s \mathcal{U}( w_*, s_* ) &=& \mathcal{J} (- i u_*).\label{Eq:U_s}
\end{eqnarray}
\label{Proposition:Derivatives}
\end{proposition}

{\bf Proof:}
For every $\ve\in\mathbb{R}$ function $v(t,\ve,s) = \mathcal{U}( w_*(t) + \ve w(t), s )$ satisfies
\begin{equation}
\df{v(t,\ve,s)}{t} = w_*(t) + \ve w(t) - i s v(t,\ve,s) - ( \overline{w}_*(t) + \ve \overline{w}(t) ) v^2(t,\ve,s).
\label{Eq:Identity:v}
\end{equation}
Differentiating this identity with respect to~$\ve$ and inserting $\ve = 0$ and $s = s_*$, we obtain
\begin{equation}
\df{v_\ve(t,0,s_*)}{t} = w(t) - ( i s_* + 2 \overline{w}_*(t) u_*(t) ) v_\ve(t,0,s_*) - \overline{w}(t) u_*^2(t).
\label{Eq:v_eps}
\end{equation}
Now, using Proposition~\ref{Proposition:Green} we solve Eq.~(\ref{Eq:v_eps})
with respect to~$v_\ve(t,0,s_*)$ and obtain formula~(\ref{Eq:U_w}).

Formula~(\ref{Eq:U_s}) is justified similarly.
We differentiate~(\ref{Eq:Identity:v}) with respect to~$s$
and solve the resulting equation using Proposition~\ref{Proposition:Green}.~\qed

\section{Self-consistency equation}
\label{Sec:SC}

Suppose that Eq.~(\ref{Eq:OA}) has a solution of the form~(\ref{Ansatz:QP})
where $a(x,t+T) = a(x,t)$ for some $T > 0$. Let us define
\begin{equation}
\omega = \fr{2\pi}{T}\qquad\mbox{and}\qquad u(x,t) = a\left( x, \fr{t}{\omega} \right),
\label{Scaling:omega:u}
\end{equation}
then the new function~$u(x,t)$ is $2\pi$-periodic with respect to~$t$ and satisfies
\begin{equation}
\omega \df{u}{t} = - i \Omega u + \fr{1}{2} e^{-i \alpha} \mathcal{G} u - \fr{1}{2} e^{i \alpha} u^2 \mathcal{G}\overline{u}.
\label{Eq:OA:u}
\end{equation}
Dividing Eq.~(\ref{Eq:OA:u}) by~$\omega$ and introducing the notations
\begin{equation}
s = \fr{\Omega}{\omega}\qquad\mbox{and}\qquad w(x,t) = \fr{e^{-i\alpha}}{2 \omega} \mathcal{G} u,
\label{Scaling:s:w}
\end{equation}
we rewrite Eq.~(\ref{Eq:OA:u}) in the form
\begin{equation}
\df{u}{t} = w(x,t) - i s u - \overline{w}(x,t) u^2.
\label{Eq:OA:u_}
\end{equation}
In Section~\ref{Sec:Riccati} we showed that every stable solution to Eq.~(\ref{Eq:OA:u_})
that lies entirely in the unit disc~$\overline{\mathbb{D}}$ is given by the formula $u(x,t) = \mathcal{U}( w(x,t), s )$.
Inserting this result into the definition of~$w(x,t)$
we arrive at a self-consistency equation
$$
w(x,t) = \fr{e^{-i\alpha}}{2 \omega} \mathcal{G} \mathcal{U}( w(x,t), s ),
$$
which can be written in the equivalent form~(\ref{Eq:SC}).

Eq.~(\ref{Eq:OA:u}) has several continuous symmetries.
More precisely, the set of its solutions is invariant with respect to the following transformations:
\smallskip

1) spatial translations $u(x,t) \mapsto u(x + c,t)$ for $c\in\mathbb{R}$,
\smallskip

2) complex phase shifts $u(x,t) \mapsto u(x,t) e^{i \phi}$ for $\phi\in\mathbb{R}$,
\smallskip

3) time shifts $u(x,t) \mapsto u(x,t+\tau)$ for $\tau\in\mathbb{R}$.
\smallskip

\noindent
All these symmetries are inherited by the self-consistency equation~(\ref{Eq:SC}),
therefore to select its unique solution~$w(x,t)$ we need to provide three pinning conditions.
In practice, this number can be reduced by one if we restrict Eq.~(\ref{Eq:SC}) to the space of even functions
$$
X_\mathrm{e} = \left\{ w \in C_\mathrm{per}([-\pi,\pi]\times[0,2\pi];\mathbb{C})\::\: w(-x,t) = w(x,t)\quad\mbox{for all}\quad (x,t)\in [-\pi,\pi]\times[0,2\pi] \vphantom{\sum}\right\}.
$$
Indeed, for symmetric coupling kernels~$G(x)$
equation~(\ref{Eq:SC}) is reflection symmetric with respect to~$x$,
therefore we can look for solutions~$w(x,t)$ satisfying $w(-x,t) = w(x,t)$ only.
In this case the spatial translation symmetry is eliminated automatically.
Then two pinning conditions relevant to the complex phase shift and the time shift can be chosen in the form
\begin{eqnarray}
&&
\Imag\left(\int_{-\pi}^\pi dx \int_0^{2\pi} w(x,t) dt \right) = 0,
\label{Eq:Pinning:1}\\[2mm]
&&
\Imag\left(\int_{-\pi}^\pi dx \int_0^{2\pi} w(x,t) e^{i t} dt \right) = 0.
\label{Eq:Pinning:2}
\end{eqnarray}
In the next sections we will show that the augmented system
consisting of Eqs.~(\ref{Eq:SC}), (\ref{Eq:Pinning:1}) and~(\ref{Eq:Pinning:2}) is well-defined.
This means that for fixed phase lag~$\alpha$ and kernel~$G(x)$ it correctly determines
the unknown even function~$w(x,t)$ and two scalar parameters~$\omega$ and~$s$.

\subsection{Modified self-consistency equation}
\label{Eq:SC:Modified}

In this section we show that the phase shift symmetry can also be eliminated from Eq.~(\ref{Eq:SC}).
Then we decrease the number of equations and unknowns in the augmented system described above.

Let us define a linear operator
$$
\mathcal{P}\::\:C_\mathrm{per}([-\pi,\pi]\times[0,2\pi];\mathbb{C})\to\mathbb{C},\qquad\mathcal{P} w = \fr{1}{(2\pi)^2} \int_{-\pi}^\pi dx \int_0^{2\pi} w(x,t) dt,
$$
which gives a constant part of the function $w(x,t)$.
Using this operator and the identity operator~$\mathcal{I}$,
we rewrite Eq.~(\ref{Eq:SC}) in the equivalent form
\begin{eqnarray}
2 \omega e^{i \alpha} \mathcal{P} w &=& \mathcal{P} \mathcal{G} \mathcal{U}(w,s),\label{Eq:Pw}\\[2mm]
2 \omega e^{i \alpha} (\mathcal{I} - \mathcal{P}) w &=& (\mathcal{I} - \mathcal{P}) \mathcal{G} \mathcal{U}(w,s).
\nonumber
\end{eqnarray}
Dividing the latter equation by the former one (which is a scalar equation!) we obtain
$$
\fr{(\mathcal{I} - \mathcal{P}) w}{\mathcal{P} w} = \fr{(\mathcal{I} - \mathcal{P}) \mathcal{G} \mathcal{U}(w,s)}{\mathcal{P} \mathcal{G} \mathcal{U}(w,s)},
$$
or equivalently
\begin{equation}
( \mathcal{P} \mathcal{G} \mathcal{U}(w,s) ) (\mathcal{I} - \mathcal{P}) w = ( \mathcal{P} w ) (\mathcal{I} - \mathcal{P}) \mathcal{G} \mathcal{U}(w,s).
\label{Eq:SC:Reduced}
\end{equation}
If we assume
\begin{equation}
w(x,t) = p + v(x,t),\quad\mbox{where}\quad p\in(0,\infty)\quad\mbox{and}\quad v(x,t)\in \left\{ u\in X_\mathrm{e}\::\: \mathcal{P} u = 0 \right\},
\label{Ansatz:w}
\end{equation}
then pinning condition~(\ref{Eq:Pinning:1}) is fulfilled automatically and can be discarded.
Moreover, inserting the ansatz~(\ref{Ansatz:w}) into Eq.~(\ref{Eq:SC:Reduced}) and into the pinning condition~(\ref{Eq:Pinning:2}) we obtain
\begin{equation}
( \mathcal{P} \mathcal{G} \mathcal{U}(p + v,s) ) v = p (\mathcal{I} - \mathcal{P}) \mathcal{G} \mathcal{U}(p + v,s),
\label{Eq:SC:Reduced_}
\end{equation}
and
\begin{equation}
\Imag\left(\int_{-\pi}^\pi dx \int_0^{2\pi} v(x,t) e^{i t} dt \right) = 0.
\label{Eq:Pinning:2_}
\end{equation}
Now instead of solving the system of equations~(\ref{Eq:SC}), (\ref{Eq:Pinning:1}) and~(\ref{Eq:Pinning:2}),
we can look for solutions of the system comprising Eqs.~(\ref{Eq:SC:Reduced_}) and~(\ref{Eq:Pinning:2_}).
In this case $p > 0$ must be given, then the system of equations~(\ref{Eq:SC:Reduced_}) and~(\ref{Eq:Pinning:2_})
has to be solved with respect to two unknowns: scalar parameter~$s$ and even function $v(x,t)$ satisfying $\mathcal{P} v = 0$.
As soon as such solution is found, one can compute the corresponding values of~$\omega$ and~$\alpha$
from Eq.~(\ref{Eq:Pw}) written in the form
$$
2 \omega e^{i \alpha} = \fr{1}{p} \mathcal{P} \mathcal{G} \mathcal{U}(p + v,s).
$$

\subsection{Modified self-consistency equation for cosine kernel~(\ref{Coupling:Cos})}
\label{Eq:SC:Modified:Cos}

In this section we consider a specific example of integral operator~$\mathcal{G}$
and show how system~(\ref{Eq:SC:Reduced_}), (\ref{Eq:Pinning:2_})
can be solved approximately using Galerkin's method.
For this we assume that~$G(x)$ is the cosine kernel~(\ref{Coupling:Cos}).

Given a positive integer~$F$ let us define a set of $8 F + 2$ functions $\psi_k(x,t)$
\begin{eqnarray*}
&& \sqrt{2} \cos x,\quad i \sqrt{2} \cos x,\\[2mm]
&& e^{i m t}, \quad i e^{i m t},\quad e^{i m t} \sqrt{2} \cos x,\quad i e^{i m t} \sqrt{2} \cos x,\quad m=-F,\dots,-1, 1,\dots,F.
\end{eqnarray*}
Note that $m\ne 0$, therefore constant functions~$\sqrt{2}$ and~$i \sqrt{2}$ are not included in the set.
The order of~$\psi_k(x,t)$ is irrelevant apart from the only place below where we will assume $\psi_8(x,t) = i e^{-i t}$.
It is easy to verify that $\psi_k(x,t)$ satisfy the orthonormality condition $\langle \psi_k , \psi_n \rangle = \delta_{kn}$
with respect to the scalar product
$$
\langle u, v \rangle = \Real\left( \fr{1}{(2\pi)^2} \int_{-\pi}^\pi dx \int_0^{2\pi} \overline{u}(x,t) v(x,t) dt\right),
$$
where~$\delta_{kn}$ is the Kronecker delta.
Therefore, functions~$\psi_k(x,t)$ span a finite-dimensional subspace of $\left\{ u\in X_\mathrm{e}\::\: \mathcal{P} u = 0 \right\}$.
We look for approximate solution to Eq.~(\ref{Eq:SC:Reduced_}) in the form
\begin{equation}
v(x,t) = \sum\limits_{k=1}^{8 F + 2} c_k \psi_k(x,t)
\label{Ansatz:v}
\end{equation}
where~$c_k\in\mathbb{R}$ are unknown coefficients.
Inserting~(\ref{Ansatz:v}) into Eq.~(\ref{Eq:SC:Reduced_}) we write $8 F + 2$ orthogonality conditions
$$
\left\langle \psi_n, ( \mathcal{P} \mathcal{G} \mathcal{U}(p + v,s) ) v \right\rangle = p \left\langle \psi_n, (\mathcal{I} - \mathcal{P}) \mathcal{G} \mathcal{U}(p + v,s) \right\rangle,\qquad n=1,\dots,8 F + 2.
$$
Since for the cosine kernel~$G(x)$ it holds $\mathcal{P} \mathcal{G} = \mathcal{P}$, the above system can be written as follows
\begin{equation}
\sum\limits_{k=1}^{8 F + 2} \left\langle \psi_n, \psi_k \mathcal{P} \mathcal{U}\left(p + \sum\limits_{m=1}^{8 F + 2} c_m \psi_m,s\right) \right\rangle c_k
= p \left\langle \psi_n, (\mathcal{I} - \mathcal{P}) \mathcal{G} \mathcal{U}\left(p + \sum\limits_{m=1}^{8 F + 2} c_m \psi_m,s\right) \right\rangle.
\label{System:Galerkin}
\end{equation}
To account for the pinning condition~(\ref{Eq:Pinning:2_}) we assume $\psi_8(x,t) = i e^{-i t}$, then
$$
\Imag\left(\int_{-\pi}^\pi dx \int_0^{2\pi} v(x,t) e^{i t} dt \right) = \Real\left(\int_{-\pi}^\pi dx \int_0^{2\pi} v(x,t) (-i) e^{i t} dt \right) = (2\pi)^2 \langle \psi_8, v \rangle.
$$
This means $c_8 = 0$. Inserting this identity into Eq.~(\ref{System:Galerkin})
we end up with a system of $8 F + 2$ nonlinear equations with respect to $8 F + 2$ real unknowns
(these are $8 F + 1$ coefficients~$c_k$ with $k\ne 8$ and the parameter~$s$).
The system~(\ref{System:Galerkin}) can be solved by Newton's method,
using a semi-analytic Jacobian expression
involving the derivative representations obtained in Section~\ref{Sec:Operator:U:Derivatives}.
Note that breathing chimera states typically have a very fine spatial structure,
therefore to approximate the integrals in~(\ref{System:Galerkin}) with the same accuracy,
one needs to use either a nonuniform grid with a moderate number of nodes in the $x$-direction,
or a uniform grid with a much larger number of nodes.
For example, all numerical results reported in Section~\ref{Sec:Example}
were obtained using a nonuniform grid with ca. $10^3$ discretization points in the $x$-direction
(the distribution of points, in this case, was $10$ to $100$ times denser
in the vicinity of the coherence-incoherence boundaries than in the other regions of the chimera state).
On a uniform grid, the same accuracy would be acheived only with at least~$10^5$ discretization points,
what would lead to extremely large computational times.

\subsection{Formulas for global order parameter and effective frequencies}
\label{Sec:Formulas:Z:Omega_eff}

Every solution~$z(x,t)$ to Eq.~(\ref{Eq:OA}) corresponds to a probability density $f(\theta,x,t)$ representing
a specific statistical equilibrium of the large system of coupled oscillators~(\ref{Eq:Oscillators}), see~\cite{Ome2013}.
More precisely, the function $f(\theta,x,t)$ yields the probability to find oscillator with phase~$\theta$ at position~$x$ at time~$t$
and has the characteristic property
\begin{equation}
\int_0^{2\pi} f(\theta,x,t) e^{i \theta} d\theta = z(x,t).
\label{Integral:z}
\end{equation}
Property~(\ref{Integral:z}) can be used to derive formulas for the global order parameter~(\ref{Def:Z})
as well as for the effective frequencies of oscillators~(\ref{Def:Omega_eff}).
Indeed, formula~(\ref{Def:Z}) determines the average of~$e^{i \theta_k(t)}$ over all oscillator indices~$k$.
For large enough~$N$ this average can also be computed as follows
\begin{equation}
Z(t) = \fr{1}{2\pi} \int_{-\pi}^\pi dx \int_0^{2\pi} f(\theta,x,t) e^{i\theta} d\theta = \fr{1}{2\pi} \int_{-\pi}^\pi z(x,t) dx.
\label{Z:Continuous}
\end{equation}
To obtain a similar formula for~$\Omega_{\mathrm{eff},k}$ we write Eq.~(\ref{Eq:Oscillators}) in the form
$$
\df{\theta_k}{t} = - \Imag\left( e^{i \alpha} \overline{W}_k(t) e^{i \theta_k(t)} \right)
\quad\mbox{where}\quad
W_k(t) = \frac{2 \pi}{N} \sum\limits_{j=1}^N G\left( \fr{2\pi (k - j)}{N} \right) e^{i \theta_j(t)}.
$$
Recall that $x_k = -\pi + 2\pi k / N$ is the spatial position of the $k$th oscillator,
therefore for infinitely large~$N$ points~$x_k$ densely fill the interval~$[-\pi,\pi]$.
In this case instead of the discrete set of functions~$W_k(t)$ we can consider
a function $W(x,t)$ depending on the continuous argument $x\in[-\pi,\pi]$.
Because of the property~(\ref{Integral:z})
we have $W(x,t) = (\mathcal{G} z)(x,t)$ and
\begin{eqnarray}
\Omega_\mathrm{eff}(x) &=& - \lim\limits_{\tau\to\infty} \fr{1}{\tau} \int_0^\tau \Imag\left( e^{i \alpha} \overline{W}(x,t) \int_0^{2\pi} f(\theta,x,t) e^{i \theta} d\theta \right) dt \nonumber\\[2mm]
&=&  - \Imag\left( e^{i \alpha} \lim\limits_{\tau\to\infty} \fr{1}{\tau} \int_0^\tau z(x,t)  (\mathcal{G} \overline{z})(x,t) dt \right).
\label{Omega_eff:Continuous}
\end{eqnarray}
Notice that formulas~(\ref{Z:Continuous}) and~(\ref{Omega_eff:Continuous}) are the continuum limit counterparts
of formulas~(\ref{Def:Z}) and~(\ref{Def:Omega_eff}) for any solution~$z(x,t)$ to Eq.~(\ref{Eq:OA}).
If the solution~$z(x,t)$ is taken in the form~(\ref{Ansatz:QP}), then we obtain the following proposition.

\begin{proposition}
Let the triple $( w(x,t), \omega, s)$ be a solution to the self-consistency equation~(\ref{Eq:SC})
and let $u(x,t) = \mathcal{U}( w(x,t), s )$ and $a(x,t) = u(x,\omega t)$.
Then the following formulas hold
$$
|Z(t)| = \fr{1}{2\pi} \left| \int_{-\pi}^\pi a(x,t) dx \right| = \fr{1}{2\pi} \left| \int_{-\pi}^\pi u(x,\omega t) dx \right|
$$
and
$$
\Omega_\mathrm{eff}(x) = - \mathrm{Im}\left( \fr{1}{T} \int_0^T e^{i\alpha} a(x,t) (\mathcal{G} \overline{a})(x,t) dt \right)
= - 2 \omega\: \mathrm{Im}\left( \fr{1}{2\pi} \int_0^{2\pi} u(x,t) \overline{w}(x,t) dt \right).
$$
\label{Proposition:Z:Omega_eff}
\end{proposition}

\subsection{Extraction of the parameters of breathing chimera states}
\label{Sec:Extraction}

Suppose that we observe a breathing chimera state in system~(\ref{Eq:Oscillators})
and we wish to extract from the observation its primary~$\Omega$ and secondary~$\omega$ frequencies
as well as the amplitude~$a(x,t)$ of the corresponding solution~(\ref{Ansatz:QP}) to Eq.~(\ref{Eq:OA}).
This can be done in the following way.

Let~$R_\mathrm{min}$ and~$R_\mathrm{max}$ be the minimal and the maximal values of~$|Z_N(t)|$,
see~(\ref{Def:Z}), over a sufficiently long observation time interval for a breathing chimera state.
If~$t_k$ are consecutive time moments such that the graph of~$|Z_N(t)|$
crosses the mid-level $( R_\mathrm{min} + R_\mathrm{max} ) / 2$ from below,
then the period~$T$ of breathing chimera state can be computed
as an average of all differences $t_k - t_{k-1}$.
On the other hand, if we compute the variation of the complex argument of~$Z_N(t)$
from the time moment~$t_{k-1}$ till~$t_k$, then the quotient
$$
\fr{1}{t_k - t_{k-1}} \int_{t_{k-1}}^{t_k} d \arg Z_N(t)
$$
yields an approximate value of the primary cyclic frequency~$\Omega$ in~(\ref{Ansatz:QP}).
Of course, the accuracy of this approximation improves
if the latter quotient is averaged over all possible indices~$k$.

\begin{remark}
Note that the above method for determining the breathing period $T$
relies on the assumption that the $|Z_N(t)|$-graph crosses
the mid-level $( R_\mathrm{min} + R_\mathrm{max} ) / 2$
only twice on the period (once from below and the other from above).
If this is not the case, then one needs to select another mid-level value
in the interval $(R_\mathrm{min},R_\mathrm{max})$,
which guarantees the two intersections condition.
\end{remark}

As soon as the period~$T$ and the primary frequency~$\Omega$ are known
we can find approximate values of the function~$a(x,t)$ in~(\ref{Ansatz:QP}) by
\begin{equation}
a(x_k,t) = \fr{1}{2 M + 1} \sum\limits_{j=k-M}^{k+M} e^{i ( \theta_j(t) - \Omega t )},
\label{Eq:a:approx}
\end{equation}
where the indices~$j$ are taken modulo~$N$,
$x_k = -\pi + 2\pi k/N$ is the scaled position of the $k$th oscillator
and $M = [\sqrt{N}/2]$ is the largest integer that does not exceed~$\sqrt{N}/2$.
Finally, using formulas~(\ref{Def:omega:s:w}) we can compute the secondary cyclic frequency~$\omega$
as well as the ratio~$s$ and the function~$w(x,t)$ appearing in the self-consistency equation~(\ref{Eq:SC}).

\begin{remark}
Note that using the method described above, we automatically obtain
the function~$a(x,t)$ and cyclic frequency~$\Omega$
which satisfy the calibration condition from Remark~\ref{Remark:Calibration}.
\end{remark}

\section{Stability analysis}
\label{Sec:Stability}

Suppose that Eq.~(\ref{Eq:OA}) has a solution of the form
\begin{equation}
z = a(x,t) e^{i \Omega t},
\label{Ansatz:0}
\end{equation}
where $a(x,t)$ is $T$-periodic with respect to its second argument.
To analyze the stability of this solution we insert the ansatz
$$
z = ( a(x,t) + v(x,t) ) e^{i \Omega t}
$$
into Eq.~(\ref{Eq:OA}) and linearize it with respect to the small perturbation $v(x,t)$.
In the result we obtain a linear equation with $T$-periodic coefficients
\begin{equation}
\df{v}{t} = - \eta(x,t) v + \fr{1}{2} e^{-i \alpha} \mathcal{G} v - \fr{1}{2} e^{i \alpha} a^2(x,t) \mathcal{G} \overline{v},
\label{Eq:Linear}
\end{equation}
where
\begin{equation}
\eta(x,t) = i \Omega + e^{i \alpha} a(x,t) \mathcal{G} \overline{a}.
\label{Def:N}
\end{equation}
Along with Eq.~(\ref{Eq:Linear}) it is convenient to consider its complex-conjugate version
$$
\df{\overline{v}}{t} = - \overline{\eta}(x,t) \overline{v}
+ \fr{1}{2} e^{i \alpha} \mathcal{G} \overline{v} - \fr{1}{2} e^{-i \alpha} \overline{a}^2(x,t) \mathcal{G} v.
$$
These two equations can be written in the operator form
\begin{equation}
\df{V}{t} = \mathcal{A}(t) V + \mathcal{B}(t) V,
\label{Eq:Operator}
\end{equation}
where $V(t) = ( v_1(t), v_2(t) )^\mathrm{T}$ is a function with values in $C_\mathrm{per}([-\pi,\pi] ; \mathbb{C}^2)$, and
$$
\mathcal{A}(t) V = \left(
\begin{array}{ccc}
- \eta(\cdot,t) & & 0 \\[2mm]
0 & & - \overline{\eta}(\cdot,t)
\end{array}
\right)
\left(
\begin{array}{c}
v_1 \\[2mm]
v_2
\end{array}
\right),
$$
and
$$
\mathcal{B}(t) V = \fr{1}{2}
\left(
\begin{array}{ccc}
  e^{-i \alpha} & & - e^{i \alpha} a^2(\cdot,t) \\[2mm]
- e^{-i \alpha} \overline{a}^2(\cdot,t) & & e^{i \alpha}
\end{array}
\right)
\left(
\begin{array}{c}
\mathcal{G} v_1 \\[2mm]
\mathcal{G} v_2
\end{array}
\right).
$$
For every fixed~$t$ the operators~$\mathcal{A}(t)$ and~$\mathcal{B}(t)$
are linear operators from $C_\mathrm{per}([-\pi,\pi] ; \mathbb{C}^2)$ into itself.
Moreover, they both depend continuously on~$t$
and thus their norms are uniformly bounded for all $t\in[0,T]$.

Our further consideration is concerned with the stability of the zero solution to Eq.~(\ref{Eq:Operator}).
Therefore, we are dealing only with the linear stability of the solution~(\ref{Ansatz:0}).
We apply the methods of qualitative analysis of differential equations in Banach spaces~\cite{DaleckiiKrein}.
Since~$\mathcal{A}(t)$ and~$\mathcal{B}(t)$ are uniformly bounded operators,
we can define an operator exponent
$$
\mathcal{E}(t) = \exp\left( \int_0^t ( \mathcal{A}(t') + \mathcal{B}(t') ) dt' \right).
$$
Then the solution of Eq.~(\ref{Eq:Operator}) with the initial condition $V(0) = V_0$
is given by the formula $V(t) = \mathcal{E}(t) V_0$.
Recalling that~$\mathcal{A}(t)$ and~$\mathcal{B}(t)$ are $T$-periodic,
we conclude~\cite[Chapter~V]{DaleckiiKrein} that the stability of the zero solution to Eq.~(\ref{Eq:Operator})
is determined by the spectrum of the {\it monodromy operator}~$\mathcal{E}(T)$.
Roughly speaking, the necessary condition for the stability of the zero solution to Eq.~(\ref{Eq:Operator})
is that the spectrum of the operator~$\mathcal{E}(T)$ lies entirely in the unit circle of the complex plane.
Otherwise, this solution is unstable.

The main problem in the application of the above stability criterion
is concerned with the fact that the monodromy operator~$\mathcal{E}(T)$
acts in an infinite-dimensional functional space.
Therefore, its spectrum consists of infinitely many points,
which can be arbitrarily distributed in the complex plane.
Below we use the explict form of the operators~$\mathcal{A}(t)$ and~$\mathcal{B}(t)$
and show the following properties of the monodromy operator~$\mathcal{E}(T)$:
\smallskip

(i) The spectrum of the operator~$\mathcal{E}(T)$ is bounded
and symmetric with respect to the real axis of the complex plane.
It consists of two qualitatively different parts:
essential spectrum~$\sigma_\mathrm{ess}$ and discrete spectrum~$\sigma_\mathrm{disc}$.
\smallskip

(ii) The essential spectrum~$\sigma_\mathrm{ess}$ is given by the formula
\begin{equation}
\sigma_\mathrm{ess} = \left\{  \exp\left( - \int_0^T \eta(x,t) dt \right) \::\: x\in[-\pi,\pi] \right\} \cup \{ \mathrm{c. c.} \}.
\label{Eq:Ess}
\end{equation}
\smallskip

(iii) The discrete spectrum~$\sigma_\mathrm{disc}$ comprises finitely many isolated eigenvalues~$\mu$,
which can be found using the formula $\mu = e^{\lambda T}$
where~$\lambda$ are roots of a characteristic equation specified below.
\smallskip

\begin{proposition}
The monodromy operator~$\mathcal{E}(T)$ can be written as a sum
\begin{equation}
\mathcal{E}(T) = \mathcal{E}_0(T) + \mathcal{K},
\label{Decomposition:E}
\end{equation}
where~$\mathcal{E}_0(T)$ is a multiplication operator of the form
$$
\mathcal{E}_0(T) = \exp\left( \int_0^T \mathcal{A}(t) dt \right),
$$
and~$\mathcal{K}$ is a compact operator from $C_\mathrm{per}([-\pi,\pi] ; \mathbb{C}^2)$ into itself.
\end{proposition}

{\bf Proof:}
Every function~$V(t)$ satisfying Eq.~(\ref{Eq:Operator})
and the initial condition $V(0) = V_0$ solves also integral equation
\begin{equation}
V(t) = \mathcal{E}_0(t) V_0 + \int_0^t \mathcal{E}_0(t) \mathcal{E}_0^{-1}(t') \mathcal{B}(t') V(t') dt',
\label{Eq:Volterra}
\end{equation}
where
$$
\mathcal{E}_0(t) = \exp\left( \int_0^t \mathcal{A}(t') dt' \right).
$$
On the other hand, every solution to Eq.~(\ref{Eq:Volterra}) can be decomposed into a sum
\begin{equation}
V(t) = \mathcal{E}_0(t) V_0 + W(t),
\label{Decomposition:V}
\end{equation}
where~$W(t)$ is a solution to the integral equation
\begin{equation}
W(t) = \int_0^t \mathcal{E}_0(t) \mathcal{E}_0^{-1}(t') \mathcal{B}(t') \mathcal{E}_0(t') V_0 dt' + \int_0^t \mathcal{E}_0(t) \mathcal{E}_0^{-1}(t') \mathcal{B}(t') W(t') dt'.
\label{Eq:Volterra_}
\end{equation}
The Volterra integral equation~(\ref{Eq:Volterra_}) has unique solution~$W(t)$ that continuously depends on the initial value~$V_0$.
Moreover, the mapping $V_0 \mapsto W(T)$ is a compact operator
from $C_\mathrm{per}([-\pi,\pi] ; \mathbb{C}^2)$ into itself
(recall the compactness of the operator~$\mathcal{G}$
involved in the definition of the operator~$\mathcal{B}(t)$).
This fact along with the formula~(\ref{Decomposition:V}) implies
that the monodromy operator~$\mathcal{E}(T)$
is the sum of the multiplication operator~$\mathcal{E}_0(T)$
and a compact operator abbreviated by~$W(T)$.~\qed

The spectrum of monodromy operator~$\mathcal{E}(T)$ consists of all numbers $\mu\in\mathbb{C}$
such that the difference operator $\mathcal{E}(T) - \mu \mathcal{I}$ is not invertible.
Because of the definition of~$\mathcal{A}(t)$ and~$\mathcal{B}(t)$
this spectrum is symmetric with respect to the real axis.
Moreover, since~$\mathcal{A}(t)$ and~$\mathcal{B}(t)$ are uniformly bounded for $t\in[0,T]$,
the monodromy operator~$\mathcal{E}(T)$ is bounded too,
and hence its spectrum lies in a circle of finite radius of the complex plane.
Other spectral properties of~$\mathcal{E}(T)$ follow from the decomposition formula~(\ref{Decomposition:E}).

Indeed, formula~(\ref{Decomposition:E}) implies~\cite{Kato} that
the essential spectrum of monodromy operator~$\mathcal{E}(T)$
coincides with the spectrum of multiplication operator~$\mathcal{E}_0(T)$.
Using the definition of~$\mathcal{A}(t)$ we obtain
$$
\mathcal{E}_0(t) = \left(
\begin{array}{ccc}
\Phi(x,t) & & 0 \\[2mm]
0 & & \overline{\Phi}(x,t)
\end{array}
\right),
$$
where
$$
\Phi(x,t) = \exp\left( - \int_0^t \eta(x,t') dt' \right).
$$
This allows us to calculate the spectrum of~$\mathcal{E}_0(T)$ explicitly
and obtain formula~(\ref{Eq:Ess}) for~$\sigma_\mathrm{ess}$.

\begin{remark}
Suppose that we consider a solution~(\ref{Ansatz:0}) to Eq.~(\ref{Eq:OA})
with the amplitude~$a(x,t)$, primary frequency~$\Omega$ and secondary frequency~$\omega$,
which satisfy the self-consistency equation~(\ref{Eq:SC}) with~$w(x,t)$ and~$s$ defined by~(\ref{Def:omega:s:w}).
Then inserting~(\ref{Scaling:omega:u}) and~(\ref{Scaling:s:w}) into~(\ref{Def:N}) we obtain
$$
\exp\left( - \int_0^T \eta(x,t) dt \right) = \exp\left( - \int_0^{2\pi} ( i s + 2 \overline{w}(x,t) u(x,t) ) dt \right)
$$
and therefore formula~(\ref{Eq:Ess}) and Proposition~\ref{Proposition:Dichotomy} imply
that every~$\mu\in\sigma_\mathrm{ess}$ lies either on the boundary of the unit circle~$|\mu| = 1$
or on the interval~$(0,1]$ of the real axis.
Hence the essential spectrum~$\sigma_\mathrm{ess}$ cannot be relevant to any instability
of the solution~(\ref{Ansatz:0}) obtained from the self-consistency equation~(\ref{Eq:SC}).
Note that, in general, Eq.~(\ref{Eq:OA}) may have solutions of the form~(\ref{Ansatz:0})
with unstable essential spectrum, but such solutions do not satisfy the self-consistency equation~(\ref{Eq:SC}).
\label{Remark:EssentialSpectrum}
\end{remark}

Formula~(\ref{Decomposition:E}) also implies
$$
\mathcal{E}(T) - \mu \mathcal{I} = ( \mathcal{E}_0(T) - \mu \mathcal{I} ) + \mathcal{K}.
$$
For every $\mu\notin\sigma_\mathrm{ess}$ the right-hand side of this formula
is a sum of the invertible operator $\mathcal{E}_0(T) - \mu \mathcal{I}$
and the compact operator~$\mathcal{K}$, hence it defines a Fredholm operator of index zero.
This means that apart from the essential spectrum~$\sigma_\mathrm{ess}$
the monodromy operator~$\mathcal{E}(T)$ can have a discrete spectrum~$\sigma_\mathrm{disc}$
consisting of eigenvalues of finite multiplicity.
Since the entire spectrum~$\sigma_\mathrm{ess}\cup\sigma_\mathrm{disc}$ is confined
in a bounded region of the complex plane, there can be at most finitely many such eigenvalues.
These eigenvalues can be computed only numerically and in the following we outline the way how this can be done.

\begin{proposition}
Let $\lambda$ be a complex number such that the equation
\begin{equation}
\df{V}{t} = \mathcal{A}(t) V + \mathcal{B}(t) V - \lambda V
\label{Eq:lambda}
\end{equation}
has a nontrivial $T$-periodic solution, then the number $\mu = e^{\lambda T}$
is an eigenvalue of the monodromy operator~$\mathcal{E}(T)$.
Conversely, for every nonzero $\mu\in\sigma_\mathrm{disc}$ there exists
a number~$\lambda\in\mathbb{C}$ such that Eq.~(\ref{Eq:lambda}) has a nontrivial $T$-periodic solution
and $\mu = e^{\lambda T}$.
\label{Proposition:Equivalence}
\end{proposition}

{\bf Proof:} Eq.~(\ref{Eq:lambda}) has a nontrivial $T$-periodic solution if and only if
$$
\exp\left( \int_0^T ( \mathcal{A}(t') + \mathcal{B}(t') - \lambda \mathcal{I} ) dt' \right)
= \exp\left( \int_0^T ( \mathcal{A}(t') + \mathcal{B}(t') ) dt' \right) e^{-\lambda T} = \mathcal{I}.
$$
This is equivalent to the identity $\mathcal{E}(T) = e^{\lambda T} \mathcal{I}$ that ends the proof.~\qed

\begin{remark}
Notice that formula $\mu = e^{\lambda T}$ is not a one-to-one relation between~$\lambda$ and~$\mu$.
For every~$\lambda\in\mathbb{C}$ it yields one value~$\mu$.
In contrast, given a nonzero~$\mu$ one obtains infinitely many corresponding values~$\lambda$,
namely $\lambda = (\log \mu + 2 \pi k i) / T$ with $k\in\mathbb{Z}$.
\label{Remark:Spectrum}
\end{remark}

\begin{proposition}
Let $\lambda$ be a complex number such that $e^{\lambda T}\notin\sigma_\mathrm{ess}$,
then for every continuous $T$-periodic function~$F(t)$ there exists a unique $T$-periodic solution of equation
$$
\df{V}{t} = ( \mathcal{A}(t) - \lambda \mathcal{I} ) V + F(t),
$$
which is given by
$$
V(t) = \int_0^T \mathcal{D}_\lambda(t,t') F(t') dt'
$$
where
$$
\mathcal{D}_\lambda(t,t') = ( \mathcal{I} - \mathcal{E}_\lambda(T) )^{-1} ( \mathcal{E}_\lambda(T) + \Theta(t - t') ( \mathcal{I} - \mathcal{E}_\lambda(T) )  ) \mathcal{E}_\lambda(t) \mathcal{E}_\lambda^{-1}(t')
$$
and
$$
\mathcal{E}_\lambda(t) = \exp\left( \int_0^t ( \mathcal{A}(t') - \lambda \mathcal{I} ) dt' \right) = \mathcal{E}_0(t) e^{-\lambda t}.
$$
\label{Proposition:Eq:Periodic}
\end{proposition}

{\bf Proof:} This assertion can be proved by analogy with~\cite[Proposition~A.1]{Ome2019}.~\qed

Proposition~\ref{Proposition:Eq:Periodic} implies that for every $\lambda\in\mathbb{C}$
such that $e^{\lambda T}\notin\sigma_\mathrm{ess}$ all $T$-periodic solutions~$V(t)$ of Eq.~(\ref{Eq:lambda})
satisfy also the integral equation
\begin{equation}
V(t) = \int_0^T \mathcal{D}_\lambda(t,t') \mathcal{B}(t') V(t') dt'.
\label{Eq:Integral:V}
\end{equation}
This fact can be used to compute the discrete spectrum~$\sigma_\mathrm{disc}$ numerically.
To this end let us choose $C > 0$ and consider Eq.~(\ref{Eq:Integral:V}) in the rectangular region
$$
\Pi = \{ \lambda\in\mathbb{C}\::\: | \Real \lambda | \le C,\: | \Imag \lambda | \le \pi / T \}.
$$
If we find all $\lambda\in\Pi$ such that Eq.~(\ref{Eq:Integral:V}) has a bounded nontrivial solution~$V(t)$,
then according to Proposition~\ref{Proposition:Equivalence} and Remark~\ref{Remark:Spectrum}
we also find all eigenvalues $\mu = e^{\lambda T}$ of the monodromy operator~$\mathcal{E}(T)$
lying in the circular region $e^{- C T} \le |\mu| \le e^{C T}$.
Since the spectrum of the monodromy operator~$\mathcal{E}(T)$ is bounded,
this ensures that for sufficiently large~$C$ we determine all eigenvalues~$\mu\in\sigma_\mathrm{disc}$
relevant to the stability of the solution~(\ref{Ansatz:0}).
Indeed, considering Eq.~(\ref{Eq:Integral:V}) for $\lambda\in\Pi$
we may overlook eigenvalues~$\mu$ in the circle $|\mu| \le e^{- C T}$.
However, all these eigenvalues satisfy $|\mu| < 1$ and therefore have no impact
on the stability of the solution~(\ref{Ansatz:0}).

\subsection{Computation of the discrete spectrum}
\label{Sec:DiscreteSpectrum}

Recalling the definitions of~$\mathcal{B}(t)$ and~$\mathcal{D}_\lambda(t,t')$
it is easy to see that Eq.~(\ref{Eq:Integral:V}) is a homogeneous Fredholm integral equation.
In general, it cannot be solved explicitly, but its solutions can be found approximately using Galerkin's method.
For this one needs to choose a set of linearly independent functions~$\varphi_k(x,t)$, $k=1,\dots,K$,
which are $2\pi$-periodic with respect to~$x$ and $T$-periodic with respect to~$t$.
Without loss of generality it can be assumed that these functions are orthonormalized with respect to the scalar product
$$
\llangle v_1, v_2 \rrangle = \fr{1}{2\pi T} \int_0^T dt \int_{-\pi}^\pi \overline{v}_1(x,t) v_2(x,t) dx
$$
such that $\llangle \varphi_j, \varphi_k \rrangle = \delta_{jk}$.
Then one looks for an approximate solution to Eq.~(\ref{Eq:Integral:V}) in the form
$$
V(t) = \sum\limits_{k=1}^K V_k \varphi_k(\cdot,t),\quad\mbox{where}\quad V_k\in\mathbb{C}^2.
$$
Inserting this ansatz into Eq.~(\ref{Eq:Integral:V}) and writing the projected problem, we obtain
$$
V_n = \sum\limits_{k=1}^K \llangle[\Bigg] \varphi_n, \int_0^T \mathcal{D}_\lambda(t,t') \mathcal{B}(t') \varphi_k(\cdot,t') dt' \rrangle[\Bigg] V_k,\qquad n=1,\dots,K.
$$
This is a system of linear algebraic equations for the $K$ two-dimensional coefficients~$V_k$.
Obviously, it has a nontrivial solution if and only if~$\lambda$ satisfies the characteristic equation
\begin{equation}
\det\left( \mathbf{M}_{2K}(\lambda) - \mathbf{I}_{2K} \right) = 0,
\label{Eq:Characteristic}
\end{equation}
where
$$
\mathbf{M}_{2K}(\lambda) = \left(
\begin{array}{ccc}
\mathbf{B}_{11}(\lambda) & \dots & \mathbf{B}_{1K}(\lambda) \\[2mm]
\vdots & \ddots & \vdots \\[2mm]
\mathbf{B}_{K1}(\lambda) & \dots & \mathbf{B}_{KK}(\lambda)
\end{array}
\right)
$$
is a block matrix with the $(2\times 2)$-matrix entries
\begin{equation}
\mathbf{B}_{nk}(\lambda) = \llangle[\Bigg] \varphi_n, \int_0^T \mathcal{D}_\lambda(t,t') \mathcal{B}(t') \varphi_k(\cdot,t') dt' \rrangle[\Bigg].
\label{Def:Matrix:B}
\end{equation}
Solving Eq.~(\ref{Eq:Characteristic}) one obtains approximate eigenvalues~$\lambda$ of Eq.~(\ref{Eq:Integral:V})
and hence the corresponding approximate eigenvalues $\mu = e^{\lambda T}$ of the monodromy operator~$\mathcal{E}(T)$.

Taking into account that the functions~$\varphi_k(x,t)$ appearing in the definition of matrix~$\mathbf{M}_{2K}(\lambda)$
must be $2\pi$-periodic with respect to~$x$ and $T$-periodic with respect to~$t$
it is convenient to choose them in the form of spatiotemporal Fourier modes.
More precisely, let $K_x$ and $K_t$ be two positive integers, then we assume
$$
\varphi_{nm}(x,t) = e^{i n x + 2 \pi i m t / T},\quad n=-K_x,\dots,K_x,\quad m=-K_t,\dots,K_t. 
$$
Thus we obtain a set of $K = (2 K_x + 1) (2 K_t + 1)$ functions such that $\llangle \varphi_{nm}, \varphi_{n'm'} \rrangle = \delta_{nn'}\delta_{mm'}$.

Notice that the functions~$\varphi_{nm}(x,t)$ have an important property.
If coupling kernel~$G(x)$ has a Fourier series representation
$$
G(x) = g_0 + \sum\limits_{k=1}^\infty 2 g_k \cos(k x)\quad\mbox{with}\quad
g_k = \fr{1}{2\pi} \int_{-\pi}^\pi G(x) e^{-i k x} dx = \fr{1}{2\pi} \int_{-\pi}^\pi G(x) \cos(k x) dx,
$$
then for all integer indices~$n$ and~$m$ we have
$$
\mathcal{G} \varphi_{nm} = 2\pi g_n \varphi_{nm}.
$$
This implies
$$
\mathcal{B}(t) \varphi_{nm} = 2\pi g_n \mathcal{B}_0(t) \varphi_{nm},\quad\mbox{where}\quad
\mathcal{B}_0(t) = \fr{1}{2}
\left(
\begin{array}{ccc}
  e^{-i \alpha} & & - e^{i \alpha} a^2(\cdot,t) \\[2mm]
- e^{-i \alpha} \overline{a}^2(\cdot,t) & & e^{i \alpha}
\end{array}
\right),
$$
therefore in the case of functions~$\varphi_{nm}(x,t)$ formula~(\ref{Def:Matrix:B}) can be written
\begin{equation}
\mathbf{B}_{nm n'm'}(\lambda) = \fr{g_{n'}}{2\pi T} \int_{-\pi}^\pi dx \int_0^T dt \int_0^T \overline{\varphi}_{nm}(x,t) \mathcal{D}_\lambda(t,t') \mathcal{B}_0(t') \varphi_{n'm'}(x,t') dt'.
\label{Def:Matrix:B_}
\end{equation}
The main advantage of the latter expression is that it does not contain any operators, but only explicitly known functions.
More precisely, the term $\mathcal{D}_\lambda(t,t')$ is a $(2\times 2)$-matrix
with entries depending on~$x$, $t$, $t'$ and~$\lambda$,
while $\mathcal{B}_0(t')$ is a $(2\times 2)$-matrix with entries depending on~$x$ and~$t'$.
Importantly, the triple integration in~(\ref{Def:Matrix:B_}) must be carried out
for each entry of the resulting product matrix separately.
As soon as all matrices $\mathbf{B}_{nm n'm'}(\lambda)$ are determined
they must be combined into the block matrix~$\mathbf{M}_{2K}(\lambda)$
and then the left-hand side of the characteristic equation~(\ref{Eq:Characteristic}) can be computed.

\begin{remark}
In Section~\ref{Sec:SC} we mentioned that Eq.~(\ref{Eq:OA:u}) has three continuous symmetries.
This implies that the monodromy operator~$\mathcal{E}(T)$ has three linearly independent
eigenfunctions corresponding to the unit eigenvalue $\mu = 1$.
Respectively the characteristic equation~(\ref{Eq:Characteristic})
has the triple root $\lambda = 0$, which is embedded in the essential spectrum.
\label{Remark:MultipleEV}
\end{remark}

\section{Example}
\label{Sec:Example}

Let us illustrate the performance of the methods developed in Sections~\ref{Sec:SC} and~\ref{Sec:Stability}.
For this we consider the breathing chimera state shown in Figure~\ref{Fig:QP}.
The primary frequency~$\Omega$ and the secondary frequency~$\omega$ of this state
can be extracted from the time trajectory of the global order parameter~$Z_N(t)$, see Section~\ref{Sec:Extraction}.
Then using formula~(\ref{Eq:a:approx}) we compute an approximate amplitude~$a(x,t)$
of the corresponding solution~(\ref{Ansatz:QP}) to Eq.~(\ref{Eq:OA}).
Inserting these results into formulas~(\ref{Scaling:omega:u}) and~(\ref{Scaling:s:w})
we obtain approximate values of the parameter~$s$ and function~$w(x,t)$.
Finally using continuous symmetries of Eq.~(\ref{Eq:OA:u}) we ensure
that~$w(x,t)$ is even with respect to~$x$ and satisfies the pinning conditions~(\ref{Eq:Pinning:1}) and~(\ref{Eq:Pinning:2}).
The obtained function~$w(x,t)$ can be represented as a Fourier series
$$
w(x,t) = \sum\limits_{k = -\infty}^\infty ( \hat{w}_{0,k} + \hat{w}_{1,k} \cos x ) e^{i k t}.
$$
Then the leading coefficients~$\hat{w}_{0,k}$, $\hat{w}_{1,k}$ with indices $k = -10,\dots, 10$
can be used as an initial guess in the Galerkin's system~(\ref{System:Galerkin}) with $F = 10$.
The latter system was solved using Newton's method up to the accuracy~$10^{-9}$.
The obtained set of coefficients~$c_k$ was transformed into function~$w(x,t)$
using formulas~(\ref{Ansatz:w}) and~(\ref{Ansatz:v}).
Then the corresponding solution~$u(x,t)$ to Eq.~(\ref{Eq:OA:u})
was computed using the operator~$\mathcal{U}$ defined in Section~\ref{Sec:Operator:U}.

\begin{figure}[ht]
\begin{center}
\includegraphics[width=0.45\textwidth,angle=0]{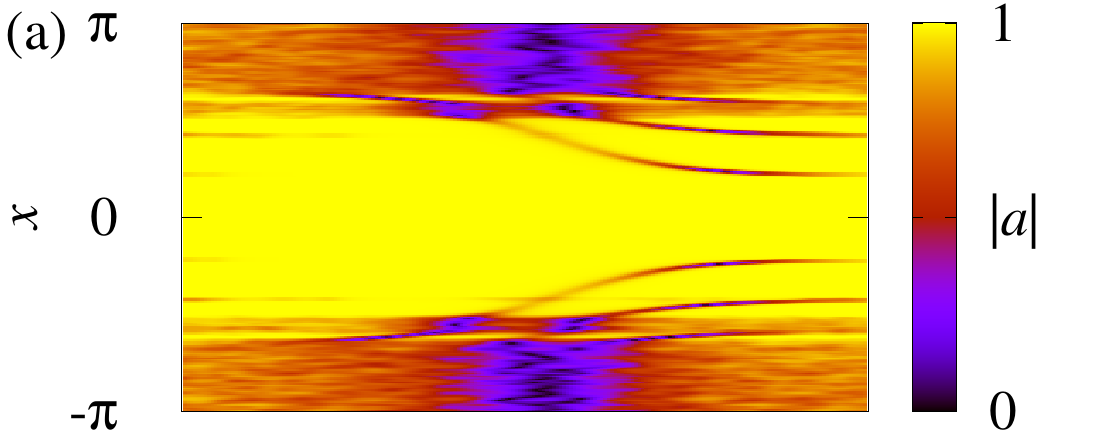}\hspace{5mm}
\includegraphics[width=0.45\textwidth,angle=0]{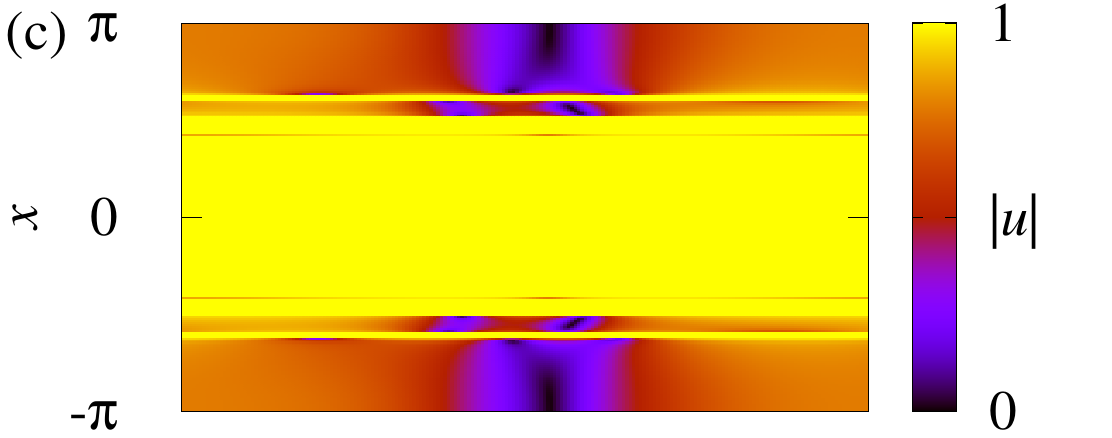}\\[2mm]
\includegraphics[width=0.45\textwidth,angle=0]{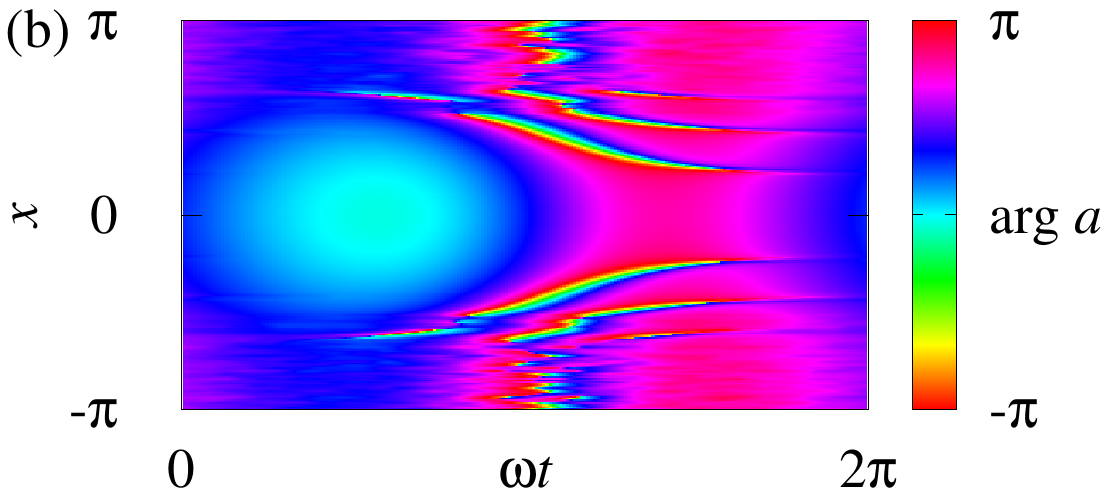}\hspace{5mm}
\includegraphics[width=0.45\textwidth,angle=0]{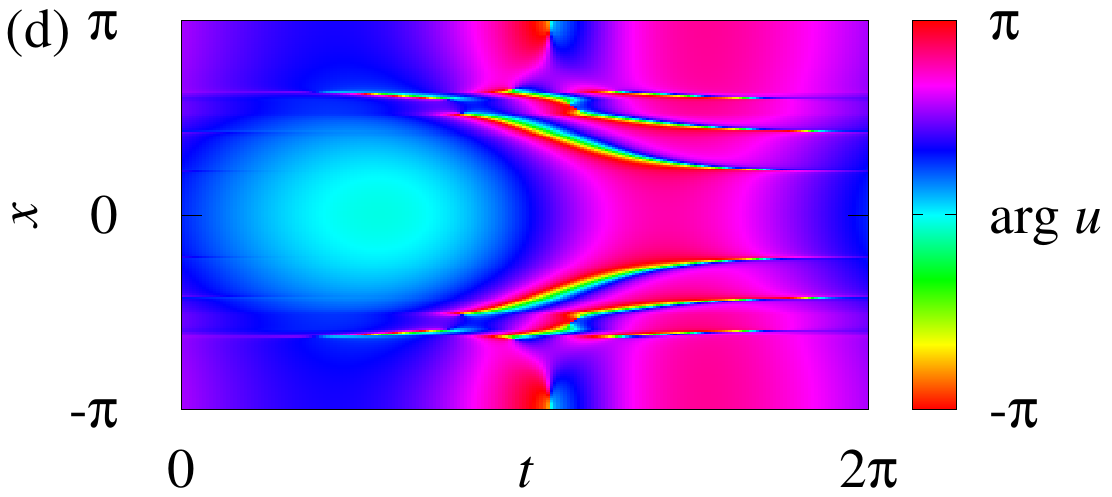}
\end{center}
\caption{(a), (b) Approximate complex amplitude~$a(x,t)$ of the solution~(\ref{Ansatz:QP})
corresponding to the chimera state in Fig.~\ref{Fig:QP}, see formula~(\ref{Eq:a:approx}).
(c), (d) The corresponding solution~$u(x,t)$ to Eq.~(\ref{Eq:OA:u})
obtained from the Galerkin's system~(\ref{System:Galerkin}) with $F = 10$.}
\label{Fig:LOP}
\end{figure}
Figure~\ref{Fig:LOP}(a),(b) shows an approximate amplitude~$a(x,t)$
computed using formula~(\ref{Eq:a:approx}) for the breathing chimera state in Fig.~\ref{Fig:QP}.
We also solved the self-consistency equation~(\ref{Eq:SC}) for the same parameters~$A$ and~$\alpha$
using Galerkin's system~(\ref{System:Galerkin}) and found a time-periodic solution~$u(x,t)$ to Eq.~(\ref{Eq:OA:u}),
see Figure~\ref{Fig:LOP}(c),(d).
As expected, the graphs of~$a(x,t)$ and~$u(x,t)$ agree with each other on a large scale,
but have some fine structure differences which can be attributed to finite size effects.
The assertion is confirmed by simulations of system~(\ref{Eq:Oscillators}) with more oscillators (not shown).
In particular, several darker filaments protruding into the coherent region (yellow/bright) in Fig.~\ref{Fig:LOP}(a)
become thinner for growing system size and disappear in the limit $N\to\infty$, see Fig.~\ref{Fig:LOP}(c),
in accordance with the coherence/incoherence invariance property described in Proposition~\ref{Proposition:Disc}.

The self-consistency equation~(\ref{Eq:SC}) allows us to predict almost perfectly
the graphs of the global order parameter~$Z_N(t)$, see~(\ref{Def:Z}),
and of the effective frequencies~$\Omega_{\mathrm{eff},k}$, see~(\ref{Def:Omega_eff}).
\begin{figure}[ht]
\begin{center}
\includegraphics[width=0.3\textwidth,angle=270]{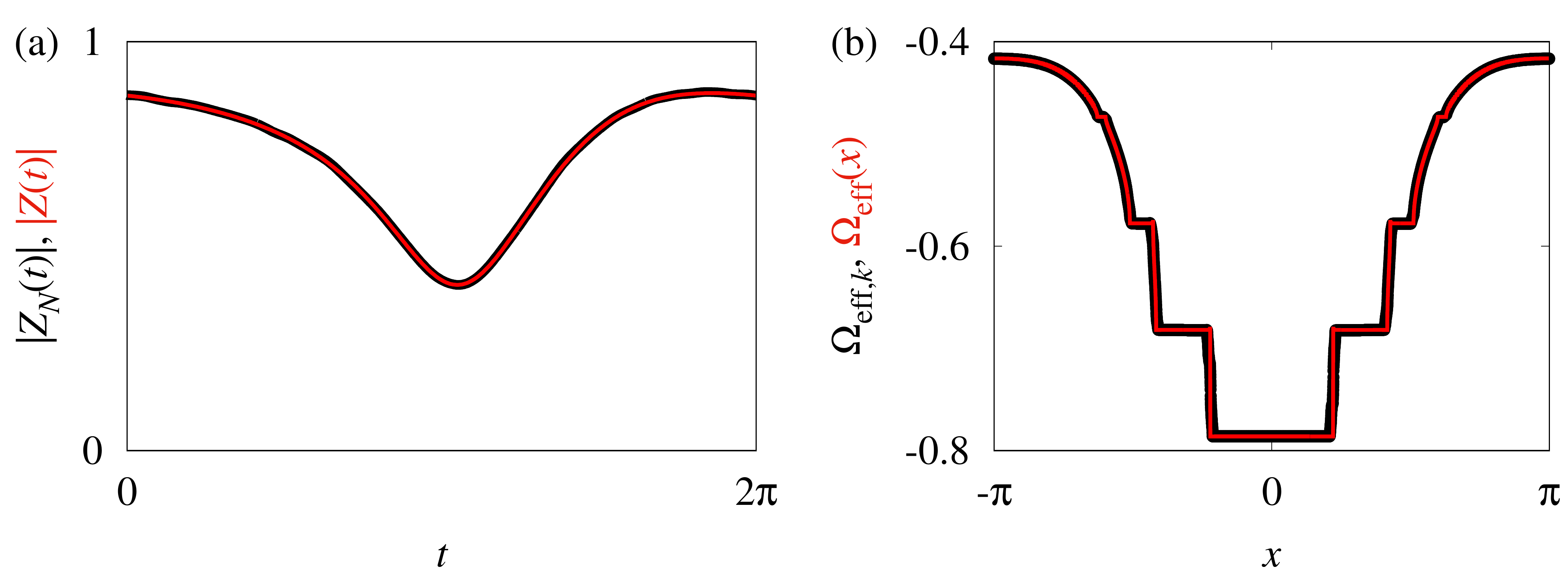}
\end{center}
\caption{(a) Global order parameter~$Z_N(t)$ and (b) effective frequencies~$\Omega_{\mathrm{eff},k}$
computed for the chimera state in Fig.~\ref{Fig:QP} and the corresponding theoretical predictions~$Z(t)$
and~$\Omega_\mathrm{eff}(x)$ obtained using Proposition~\ref{Proposition:Z:Omega_eff}.}
\label{Fig:Comparison}
\end{figure}
In Figure~\ref{Fig:Comparison} these quantities, computed
for a breathing chimera state in system~(\ref{Eq:Oscillators}),
are compared with their continuum limit counterparts~$Z(t)$ and~$\Omega_\mathrm{eff}(x)$
computed by the formulas from Proposition~\ref{Proposition:Z:Omega_eff}
where we inserted the functions~$w(x,t)$ and~$u(x,t)$ obtained from the Galerkin's system~(\ref{System:Galerkin}).

\begin{figure}[ht]
\begin{center}
\includegraphics[width=0.3\textwidth,angle=270]{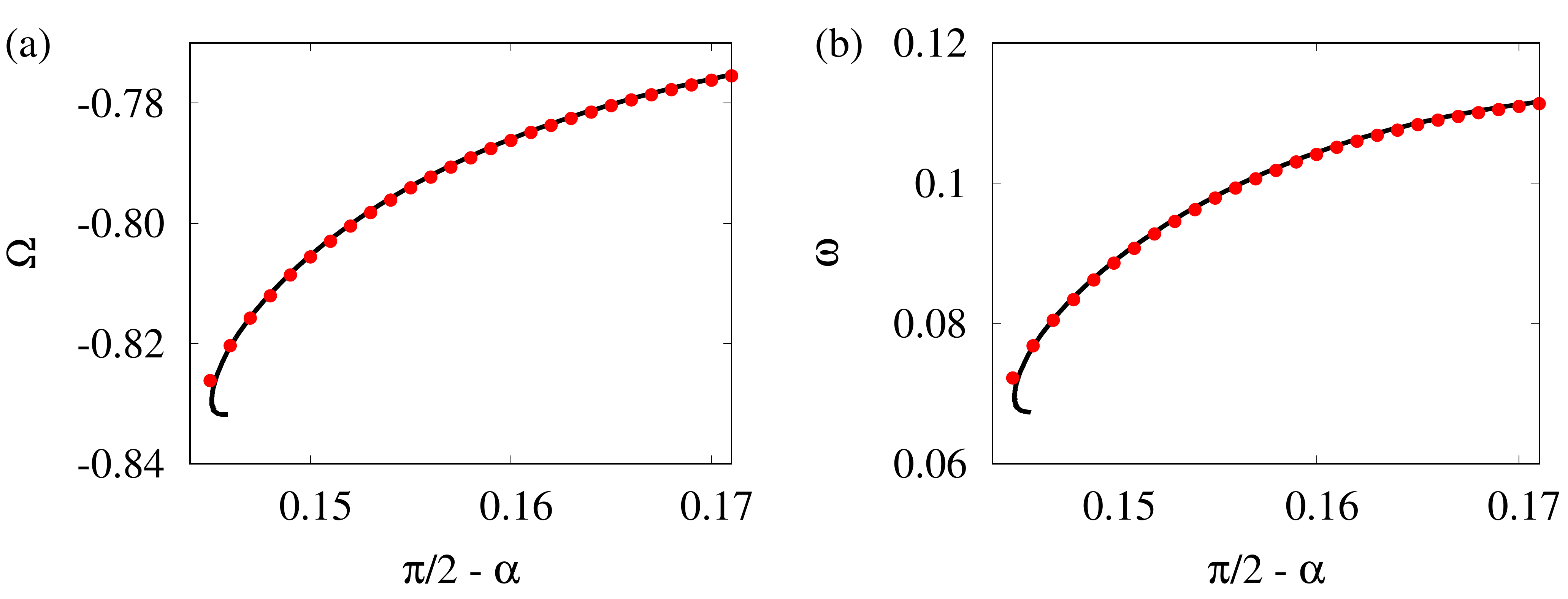}
\end{center}
\caption{(a) Primary frequency~$\Omega$ and (b) secondary frequency~$\omega$ of breathing chimera states
in the system~(\ref{Eq:Oscillators}) for cosine coupling kernel~(\ref{Coupling:Cos}) and $A = 1.05$.
The solid curve shows theoretical predictions made using the Galerkin's system~(\ref{System:Galerkin}) with $F = 10$.
The points show frequencies extracted from the breathing chimera states
observed in the system~(\ref{Eq:Oscillators}) with $N = 8192$ oscillators.}
\label{Fig:Scan}
\end{figure}

Figure~\ref{Fig:Scan} illustrates another application of the self-consistency equation~(\ref{Eq:SC}).
We used it for computation of a branch of breathing chimera states in Eq.~(\ref{Eq:OA}).
The theoretically predicted primary and secondary frequencies are compared
with the corresponding values of~$\Omega$ and~$\omega$ observed in breathing chimera states
in the coupled oscillator system~(\ref{Eq:Oscillators}) with $N = 8192$.
Again, the agreement between the theoretical curve and the numerical points is very good.
A slightly recognizable mismatch can be attributed to finite size effects
or to the small number of Fourier modes ($F = 10$) in the Galerkin's approximation.
Note that the curves in Figure~\ref{Fig:Scan} fold for $\alpha \approx \pi/2 - 0.145$.
This fact explains a sudden collapse of breathing chimera states to the completely synchronous state,
which we observed in system~(\ref{Eq:Oscillators}) for $\alpha > \pi/2 - 0.145$.

For all breathing chimera states on the solution branch in Figure~\ref{Fig:Scan}
we also computed the corresponding essential spectra~$\sigma_\mathrm{ess}$.
\begin{figure}[ht]
\begin{center}
\includegraphics[width=0.25\textwidth,angle=270]{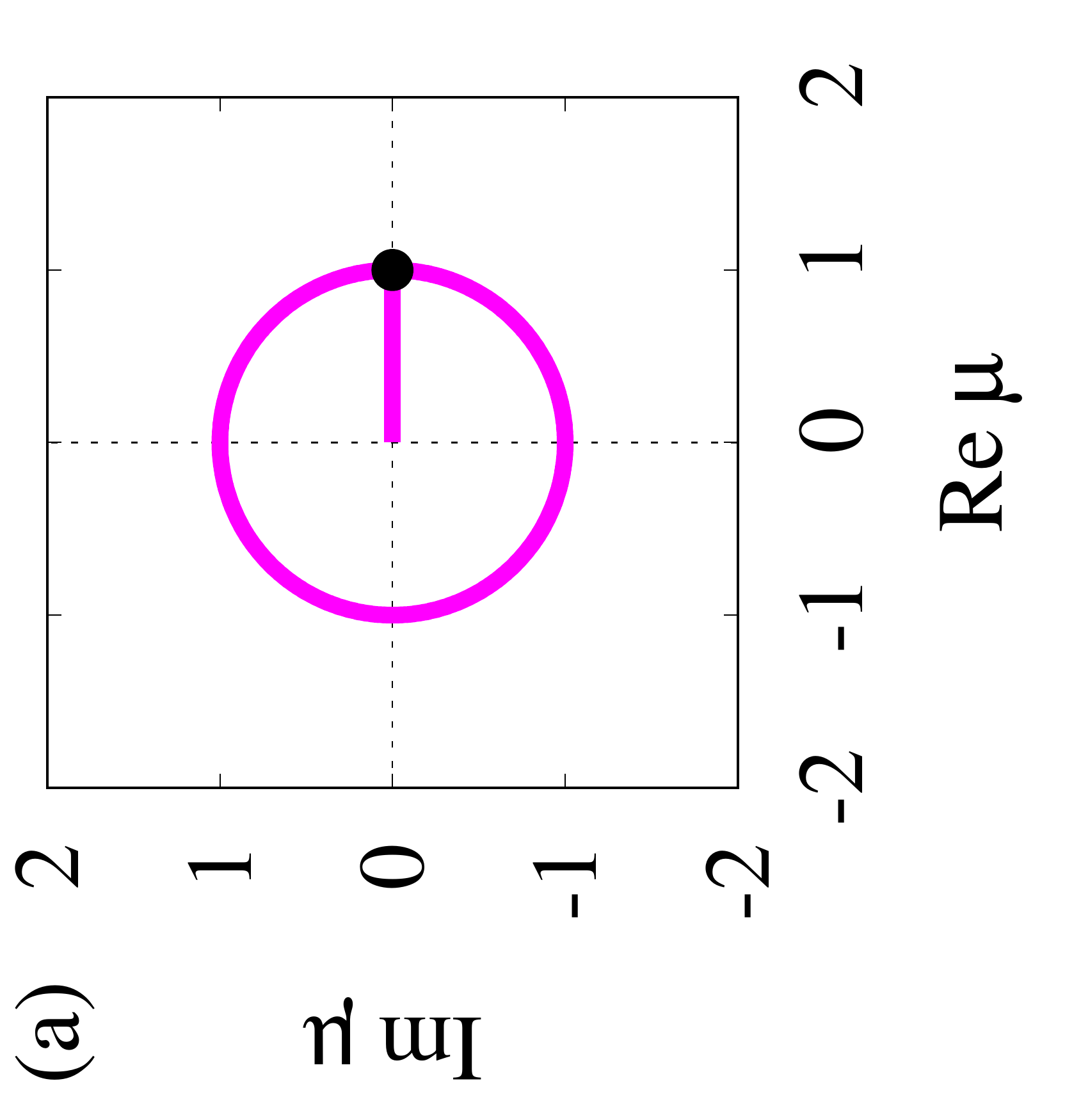}%
\includegraphics[width=0.25\textwidth,angle=270]{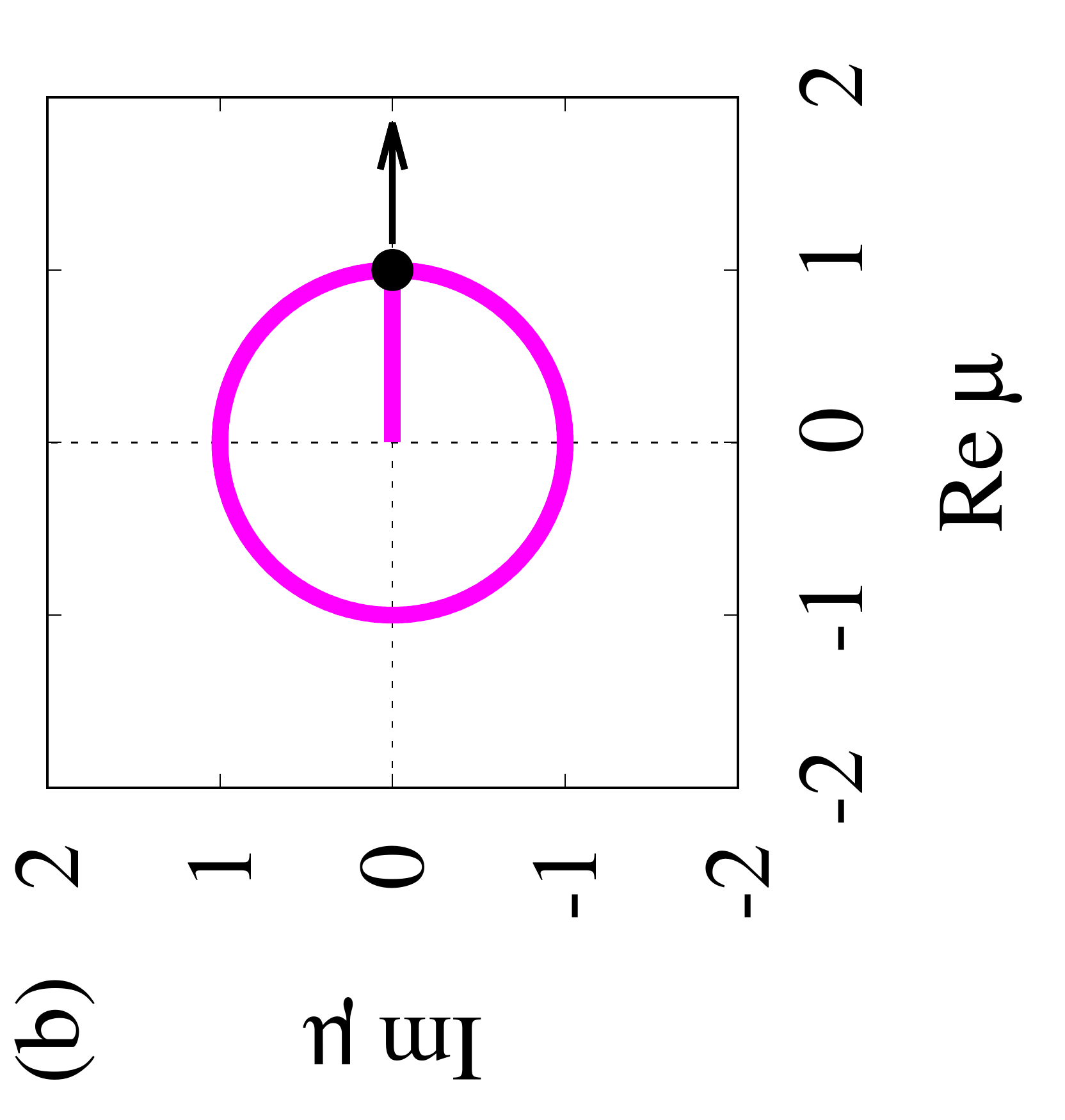}%
\includegraphics[width=0.25\textwidth,angle=270]{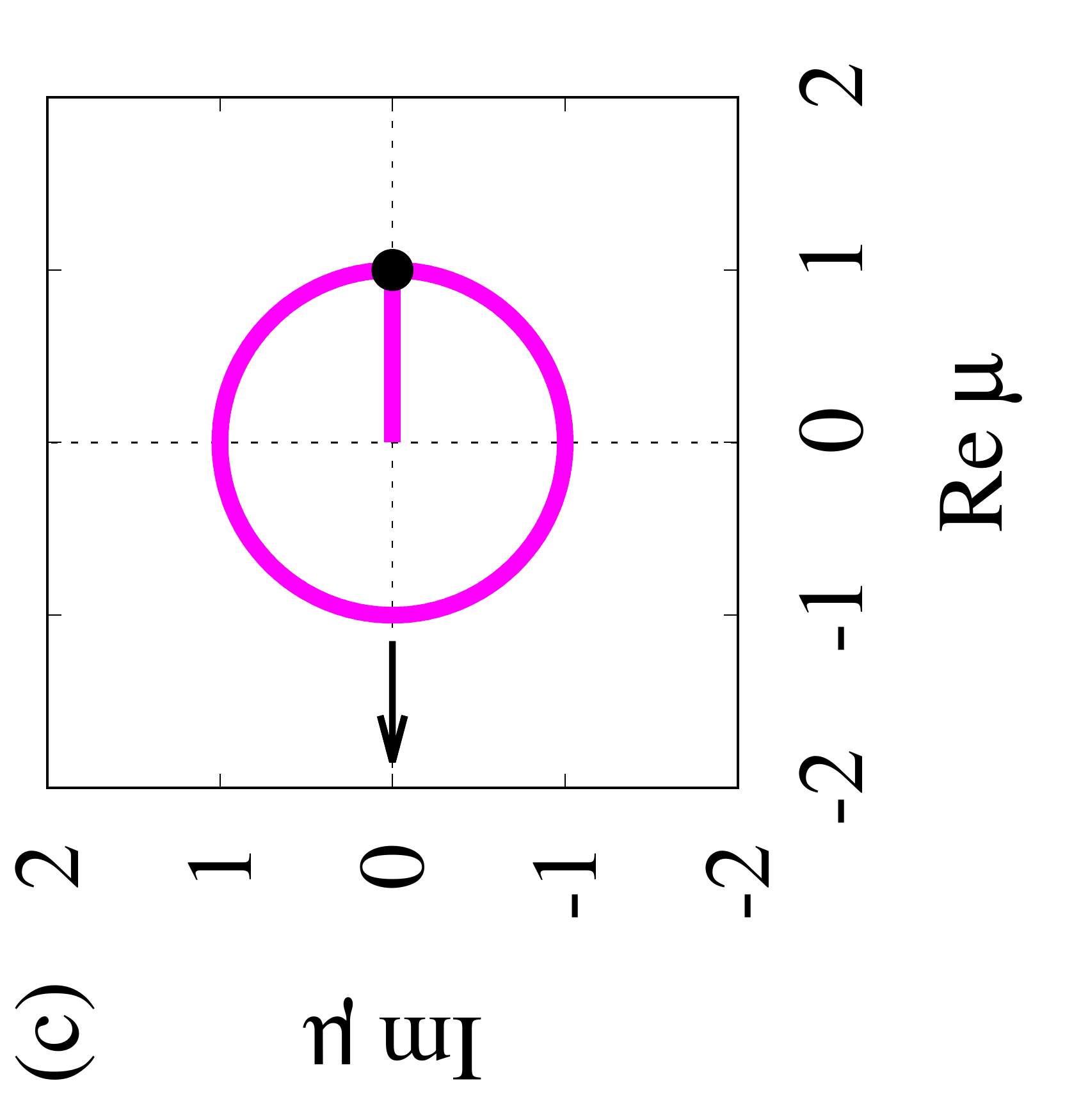}%
\includegraphics[width=0.25\textwidth,angle=270]{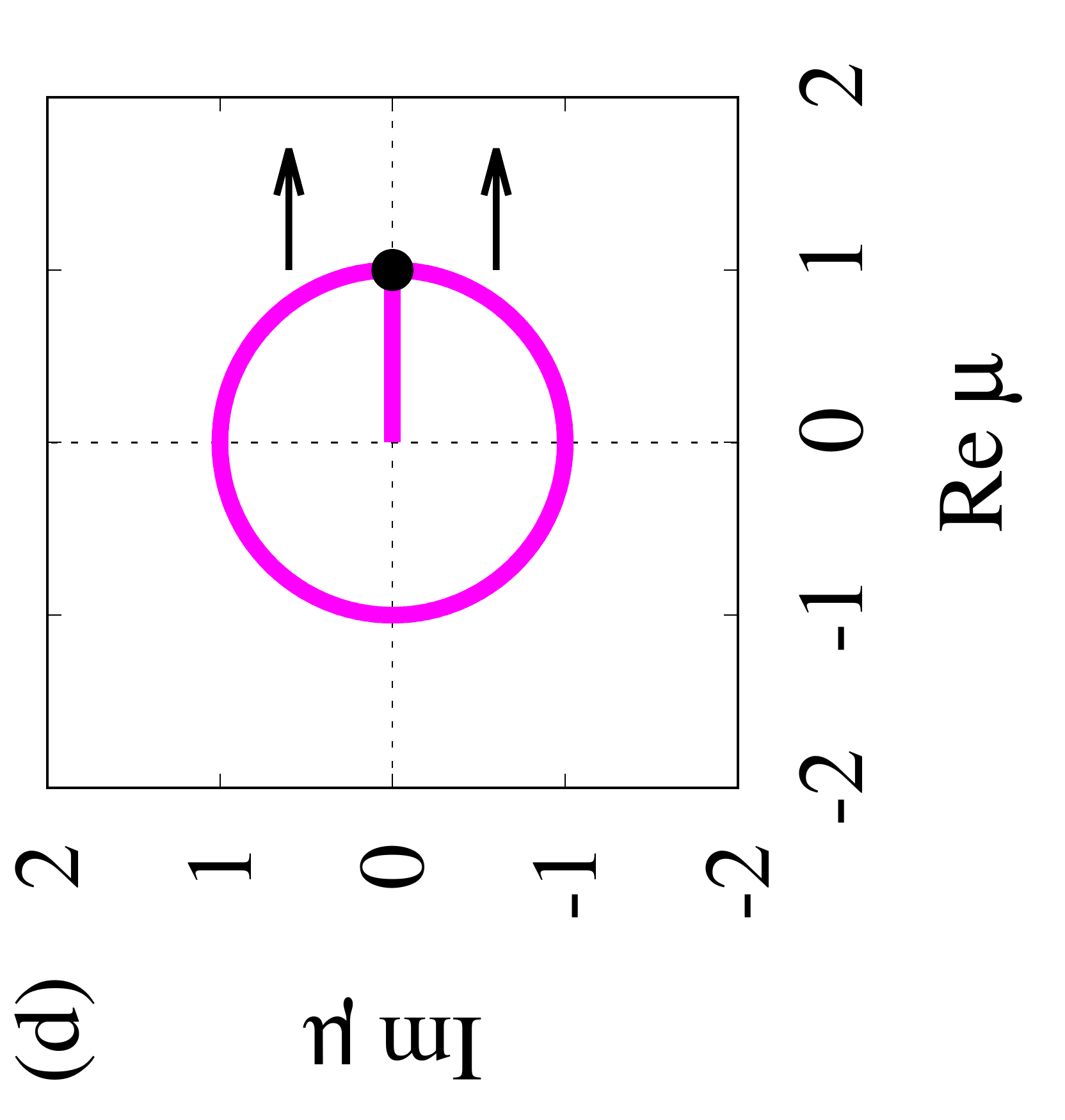}
\end{center}
\caption{All breathing chimera states on the solution branch in Fig.~\ref{Fig:Scan}
have identical essential spectra shown in panel~(a).
The point indicates a multiple eigenvalue embedded into the essential spectrum.
Other panels show hypothetical bifurcation scenarios for breathing chimera states:
(b) Fold and symmetry-breaking bifurcations, (c) period-doubling bifurcation, (d) torus bifurcation.
The arrows indicate directions in which one or two eigenvalues can appear from the essential spectrum.}
\label{Fig:Spectra}
\end{figure}
These spectra look identically, see Figure~\ref{Fig:Spectra}(a),
and have the maximal possible size, see Remark~\ref{Remark:EssentialSpectrum} for more detail.
The computation of the discrete spectra~$\sigma_\mathrm{disc}$ turned out
to be a time demanding task, therefore at present we were unable to carry it out.
However, because of Remark~\ref{Remark:MultipleEV} we assert that~$\sigma_\mathrm{disc}$
includes a triple eigenvalue $\mu = 1$ embedded into the essential spectrum.
Since for every breathing chimera state the unit circle~$|\mu| = 1$
is a subset of its essential spectrum~$\sigma_\mathrm{ess}$,
the destabilization of such chimera state cannot occur via a classical bifurcation of finite codimension.
Indeed, any unstable eigenvalue can emerge only from the essential spectrum
and therefore this eigenvalue cannot be isolated at the bifurcation point.
By analogy with other dynamical systems we may expect
that breathing chimera states in general can lose their stability
via a fold, symmetry breaking, period-doubling or torus bifurcation,
which are associated with the appearance of one or two unstable eigenvalues~$\mu$
from the essential spectrum on the unit circle as shown in Figure~\ref{Fig:Spectra}(b)-(d).
Note that proper consideration of such bifurcations requires the use of generalized spectral methods~\cite{ChiN2011,Die2016,ChiM2019},
which, however, must be adapted to a situation where the reference solution is a relative periodic orbit rather than a simple equilibrium.

\section{Conclusions}
\label{Sec:Conclusions}

In summary, we showed that breathing chimera states observed in large size systems~(\ref{Eq:Oscillators})
are properly represented by solutions of the form~(\ref{Ansatz:QP}) in Eq.~(\ref{Eq:OA}).
The self-consistency equation~(\ref{Eq:SC}) and Proposition~\ref{Proposition:Z:Omega_eff}
allow one to predict the most important features of such chimera states.
A general approach for stability analysis of breathing chimera states is formulated in Section~\ref{Sec:Stability}.
It relies on the consideration of the monodromy operator which describes the evolution of small perturbations in the system.
We found that the spectrum of this operator consists of two qualitatively different parts: essential and discrete spectra.
The former part was completely characterized in this paper,
while for the latter part we suggested a numerical algorithm for its approximate evaluation.
Although we were unable to compute discrete spectrum of any breathing chimera state
we obtained a theoretical indication of the fact that such chimera states lose their stability
in nonclassical bifurcation scenarios where one or two unstable eigenvalues
appear from the essential spectrum. Note that this statement is still speculative
and therefore needs to be confirmed by computable examples.

We emphasize that the consideration scheme suggested in this paper can be applied to systems~(\ref{Eq:Oscillators})
with arbitrary coupling kernels~$G(x)$, including exponential~\cite{BolSOP2018}
and top-hat~\cite{SudO2018,SudO2020} coupling.
In particular, using the self-consistency equation~(\ref{Eq:SC})
one can carry out a more rigorous asymptotic analysis of breathing chimera states
reminiscent of that in~\cite{SudO2020}.
Moreover, all above methods can be extended to the  study of breathing spiral chimera states
in two-dimensional lattices of phase oscillators~\cite{XieKK2015,OmeWK2018}.
Furthermore, our results can also be applied to explore the appearance
of pulsing and alternating coherence-incoherence patterns~\cite{Ome2020a}
and modulated travelling chimera states~\cite{Ome2020}
in systems of heterogeneous nonlocally coupled phase oscillators,
though in this case one needs to modify the definition of the solution operator~$\mathcal{U}$.

In a more general context, we believe that many theoretical constructions of this paper
such as the solution operator of the periodic Riccati equation~(\ref{Eq:Riccati}),
the periodic self-consistency equation~(\ref{Eq:SC})
and the essential and discrete spectra of the monodromy operator~$\mathcal{E}(T)$
can be adapted to a broader class of problems
concerned with the application of the Ott-Antonsen ansatz~\cite{OttA2008}.
These are, for example, the Kuramoto model with a bimodal frequency distribution,
where one finds periodically breathing partially synchronized states~\cite{MarBSOSA2009},
modular Kuramoto networks~\cite{BicGLM2020} as well as networks of theta neurons~\cite{LukBS2013,Lai2014,ByrAC2019}
(or equivalently, quadratic integrate-and-fire neurons~\cite{MonPR2015,Esn-ARAM2017}).

\section*{Acknowledgment}

This work was supported by the Deutsche Forschungsgemeinschaft under grant OM 99/2-1.

\end{document}